\documentclass[12pt,preprint,longabstract]{aastex}

\usepackage{hyperref}
\usepackage{lscape}
\usepackage{multirow}
\usepackage{natbib}

\shorttitle{Physical Parameters of KIC 6131659}
\shortauthors{Bass et al.}

\def\fss{\hbox{$.\!\!^s$}}

\begin{document}

\title{{\em Kepler} Studies of Low-Mass Eclipsing Binaries I. Parameters of the Long-Period Binary KIC 6131659$^{\dagger}$}

\author{Gideon Bass\altaffilmark{1}, Jerome A. Orosz\altaffilmark{1}, William F. Welsh\altaffilmark{1}, Gur Windmiller\altaffilmark{1}, Trevor Ames Gregg\altaffilmark{1}, Tara Fetherolf\altaffilmark{1}, Richard A. Wade\altaffilmark{2}, Samuel N. Quinn\altaffilmark{3}}

\altaffiltext{1}{Department of Astronomy, San Diego State University, 5500 Campanile Dr. San Diego, CA 92182 and School of Physics, Astronomy and Computational Sciences, George Mason University 4400 University Dr, Fairfax, VA 22030}
\altaffiltext{2}{Department of Astronomy \& Astrophysics, The Pennsylvania State University, 525 Davey Lab, University Park, PA 16802}
\altaffiltext{3}{Department of Physics \& Astronomy, Georgia State University, PO Box 4106, Atlanta, GA 30302} 
\altaffiltext{$\dagger$}{Based on observations obtained with the Hobby-Eberly Telescope, which is a joint project of the University of Texas at Austin, the Pennsylvania State University, Stanford University, Ludwig-Maximilians-Universit\"at M\"unchen, and Georg-August-Universit\"at G\"ottingen.}

\begin{abstract}
KIC 6131659
is a long-period (17.5 days) eclipsing binary discovered
by the {\em Kepler} mission.
We analyzed six quarters of
{\em Kepler} data along with supporting ground-based
photometric and spectroscopic data to obtain accurate values for the
mass and radius of both stars, namely 
$M_1=0.922\pm 0.007\,M_{\odot}$,
$R_1=0.8800\pm 0.0028\,R_{\odot}$, and 
$M_2=0.685\pm 0.005\,M_{\odot}$,
$R_2=0.6395\pm 0.0061\,R_{\odot}$.  
There is a well-known issue with low mass 
($M\lesssim 0.8\,M_{\odot}$) stars (in
cases where the mass and radius measurement uncertainties
are smaller than
two or three percent) where the measured radii are almost
always 5 to 15 percent 
larger than expected from evolutionary models,
i.e.\ the measured
radii are all above the model isochrones in a mass-radius plane.
In contrast,
the two stars in KIC
6131659 were found to sit on the same theoretical isochrone in the mass-radius plane.
Until recently, all of the well-studied eclipsing binaries with low-mass
stars had periods less than about three days.  The stars in such systems
may have been inflated by high levels of stellar activity induced by
tidal effects in these close binaries.
KIC 6131659 shows essentially no evidence of enhanced stellar activity,
and our
measurements  support the hypothesis that
the unusual mass-radius relationship observed in most
low-mass stars is
influenced
by strong magnetic activity created by the rapid rotation of the stars in
tidally-locked, short-period systems.  
Finally, using short cadence data, we show that KIC 6131657
has one of the smallest measured non-zero eccentricities of
a binary with two main sequence stars, where 
$e\cos\omega=(4.57\pm 0.02)\times 10^{-5}$.
\end{abstract}
\keywords{stars: binary: eclipsing, stars: fundamental 
parameters, stars: low-mass}

\section{Introduction}
\label{sec:Introduction}
Double-lined eclipsing binaries (DLEBs) are the best known way of
accurately measuring the masses and radii of stars
\citep[see][for a recent review]{Torres_2010}. 
Although recent discoveries in interferometry,
asteroseismology, and other methods provide promising new ways for
measuring stellar radii, these techniques still lack the precision of
DLEBs in almost all cases. 

Comparisons of DLEBs with predicted mass-radii
relations from stellar models have been made for a variety of stellar
masses. However, until recently, only a few systems in the low-mass regime 
were well characterized, mostly because of difficulties in
observing these systems. Those low-mass
systems that had been observed
showed radii consistently about 10-15\% higher than predictions from
stellar and evolutionary models \citep[see, for example][]{Torres_Ribas_2002,
Ribas_2006, LopezMorales_2007, 
Ribas_2008}.
The most common explanation for this discrepancy 
is that tidal
interactions in short period systems causes higher than normal
rotational
speeds, which in turn increase the level of stellar activity
\citep{Pizzolato_2003,LopezMorales_2007}. 

If the above explanation for the bloated radii of the low-mass stars
is true, one would expect to find low mass
binaries that have
large separations (and periods) to have a normal
mass-radius relationship. Unfortunately, it can be 
difficult to identify and observe long period DLEBs with ground based
instrumentation. K and M dwarfs can be faint, and large numbers of them
need to be monitored for long periods of time  to find eclipses.
Not that long ago, only two low mass DLEBs were known, CM Dra
\citep{Lacy_1977, Metcalfe_1996}, and YY Gem \citep{Bopp_1974,
Leung_Schneider_1978}. Recently, in the past 
decade or so, various ground based
surveys have detected several more \citep[see e.g.][]{Ribas_2006,
Cakirli_2010,Bhatti_2010,Becker_2011,
Kraus_2011}.
However, nearly all of these have had short periods
of less than about
three days.  Notable exceptions include T-Lyr1-17236 with a pair of
M-stars in an 8.4 day orbit \citep{Devor_2008},
Kepler-16 with a pair of low-mass stars in a 41 day orbit
\citep{Doyle_2011},
and LSPM J1112+7626
with a pair of M-stars that are also in a 41 day orbit \citep{Irwin_2011}.

NASA's {\em Kepler} mission offers an exciting opportunity in this
field of study.  {\em Kepler} provides a virtually uninterrupted look
at roughly 156,000 stars at an unprecedented level of sensitivity. 
\citet{Prsa_2011} identified 1832 eclipsing
binaries in the {\em
  Kepler} field, of which 95 have been identified from their broad-band
colors by \citet{Coughlin_2011}
as low mass (both
stars are
less than $\approx 1\,M_\odot$),
well-separated systems with deep eclipses
suited for ground-based follow-up.
We have obtained ground-based photometry and
spectroscopy of one of these candidates, KIC 6131659,
which is a long-period binary ($P=17.5$ d), 
and have found that both of
the stars in KIC 6131659 are not bloated and
fit nicely on the same theoretical mass-radius relation.  This
offers support for the hypothesis that the previous observational disparity
from theory was caused by tidally induced magnetic activity.

In
\S\ref{sec:Observations} we discuss the observations, and 
in \S\ref{sec:DataAnalysis} we discuss our analysis of
the data using our ELC code to model and measure
physical parameters from the observations. 
In \S\ref{results} we present results,
including calculated physical parameters, 
and explore how these results compare
with theoretical predictions. Finally, \S\ref{sec:Conclusion}
provides a short summary.

\section{Observations}
\label{sec:Observations}

\subsection{{\em Kepler} Photometry}

Discussions of the details of the {\em Kepler}\, mission are found
in \citet{Borucki_2010}, \citet{Koch_2010},
\citet{Batalha_2010},
\citet{Caldwell_2010}, and
\citet{Gilliland_2010}.
The {\em Kepler} spacecraft is in an Earth-trailing heliocentric orbit,
which allows for near continuous coverage of its 105 deg$^2$
field of view.  Many eclipsing binaries have been
discovered
among the {\em Kepler} targets  
\citep{Prsa_2011,Slawson_2011,Coughlin_2011},
including many with periods
longer than 10 days.  We initially
selected a sample of 10 binaries for ground-based followup observations,
among them was KIC 6131659 (2MASS J19370697+4126128, 
$\alpha=19^{\rm h}37^{\rm m}06\fss 98$,      
$\delta=+41^{\circ}26^{\prime}12\farcs 8$, J2000). 
It is relatively bright
($r=12.5$),
has a long period
(17.52 days), deep primary and secondary
eclipses (35\% and 10\%, respectively),
and has
out-of-eclipse variability at the $\lesssim 0.2\%$ level, 
which is a sign that
there is little to no star-spot activity.  
For this project
we obtained
the public {\em Kepler} light curves
from the MAST archive   
(Kepler quarters\footnote{The {\em Kepler}
observations are divided up into Quarters of
$\approx 90$ days each, and will be denoted by Q2 for
Quarter 2, Q3 for Quarter 3, etc.}
Q0-Q6), 
which include 28 primary eclipses and
24 secondary eclipses.

\begin{figure}
\epsscale{0.9}
\includegraphics[angle=-90,scale=0.7]{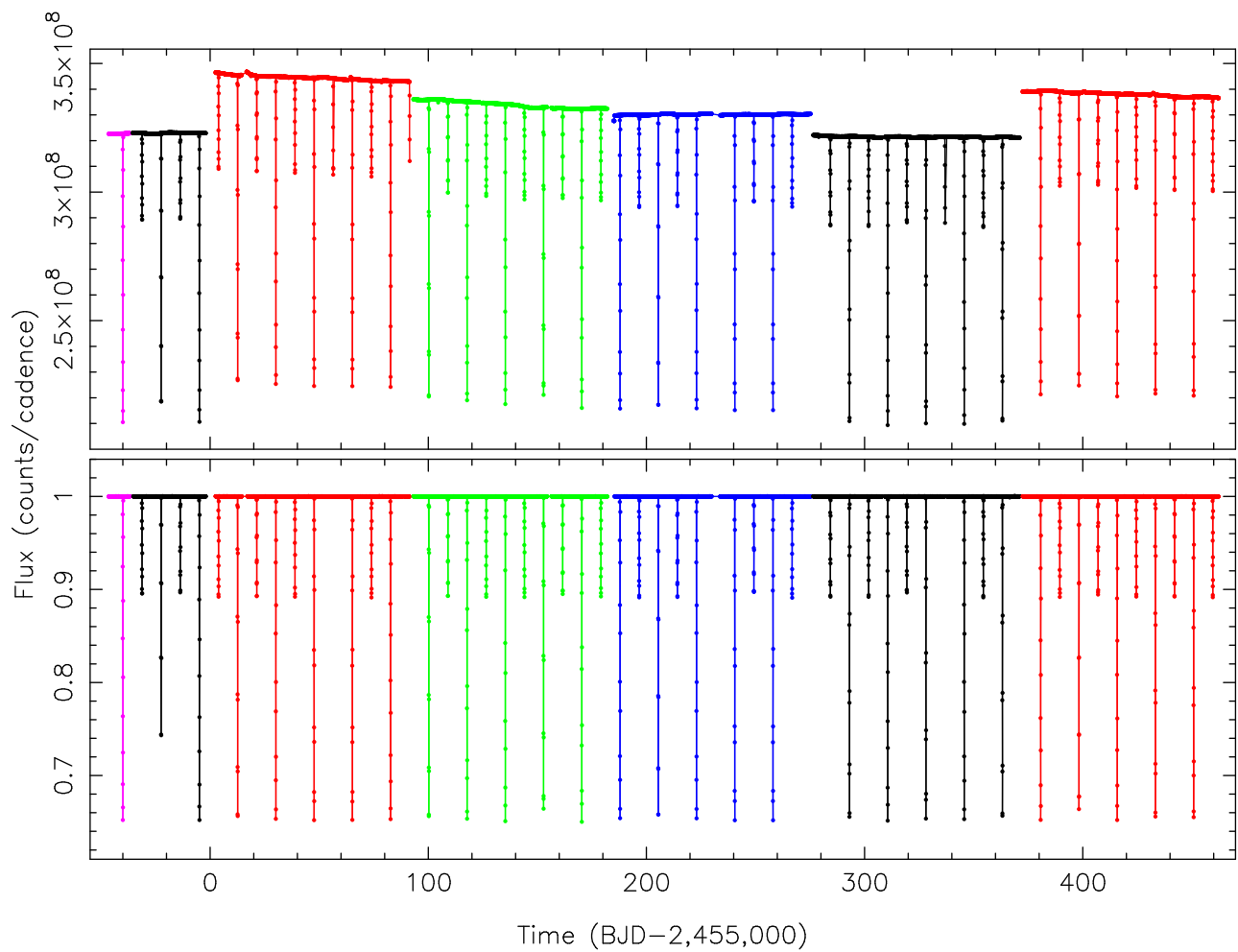}
\caption{Top:  The SAP light curve of KIC 6131659
from Q0 to Q6.  Apart from the magenta color for Q0,
the colors indicate the season 
and hence the spacecraft orientation with black for Q1 and Q5,
red for Q2 and Q6, green for Q3, and blue for Q4.
Bottom: The detrended and normalized light curve.
The first primary eclipse in Q1 has an outlier that was given a very
low weight.  One primary eclipse was missed and
a total of six secondary eclipses were only partially covered
or missed completely.
}
\label{plotraw}
\end{figure}

The {\em Kepler} data processing pipeline
\citep{Jenkins_2010a,Jenkins_2010b} outputs two types of light curves,
simple aperture photometry (SAP) light curves and ``pre-search data
conditioned'' (PDC) light curves, which have some of the instrumental
trends removed \citep{Smith_2012,Stumpe_2012}.  While recent versions
of the PDC light curves have better detrending, they still have
problems removing sudden jumps in the light curves (usually due to
cosmic rays). We have also found that the PDC process tends to
overcorrect when removing light from contaminating sources. We
therefore used the SAP data in this study, and implemented our own
detrending algorithm.  Each quarter of data is detrended
separately.
Discontinuities due to cosmic ray hits
and gaps in the light curves (usually caused by either monthly data
downloads and quarterly rolls) were identified by a visual
inspection. The light curves were then divided up into pieces using the
discontinuities and gaps as end points.  Each piece of the light curve
was normalized in a way similar to the way an optical spectrum is
normalized to its local continuum.  
The relatively flat out-of
eclipse areas of the light are treated as the
``continuum'', and the
eclipses are treated
as ``absorption lines''.  Using the fitting tools in the
IRAF\footnote{IRAF is distributed by the National Optical Astronomy
  Observatory, which is operated by the Association of Universities
  for Research in Astronomy (AURA) under cooperative agreement with
  the National Science Foundation.}  task 'splot', an $n$-piece cubic
spline interpolating function (where $n$ was
typically between 10 and 30) was
fit to each piece after the eclipses were masked out using an
iterative sigma-clipper.  Each light curve piece was divided by the
interpolating function, and the normalized pieces were reassembled to
make the complete and normalized light curve.  The SAP and detrended
light curves are shown in Figure \ref{plotraw}.

There is a single point in the middle of the first primary eclipse seen in Q1 that is 
much brighter than it should be, based on the other primary eclipses
seen. This outlier is believed to be a result of an error in the cosmic
ray rejection routine (J. Jenkins 2011, private communication).
A few other similar, but less deviant, outliers were identified
in other eclipses. These points were given a very low weight in the analysis described below.  
Also, although the duty cycle is very high, the light curve is not complete.
A secondary eclipse was in progress during a monthly data download
in Q2, and one was in progress when the  spacecraft performed its quarterly roll at the end of Q2. Likewise in Q5, a secondary
eclipse was interrupted by a monthly data download. Since incomplete coverage may introduce errors
in the detrending, we excluded all three of these partially covered events entirely.
A complete primary eclipse was missed during the spacecraft roll after Q4, and secondary eclipses were
missed entirely during the spacecraft rolls following Q2 and Q5.

\begin{figure}
\epsscale{0.9}
\includegraphics[angle=-90,scale=0.7]{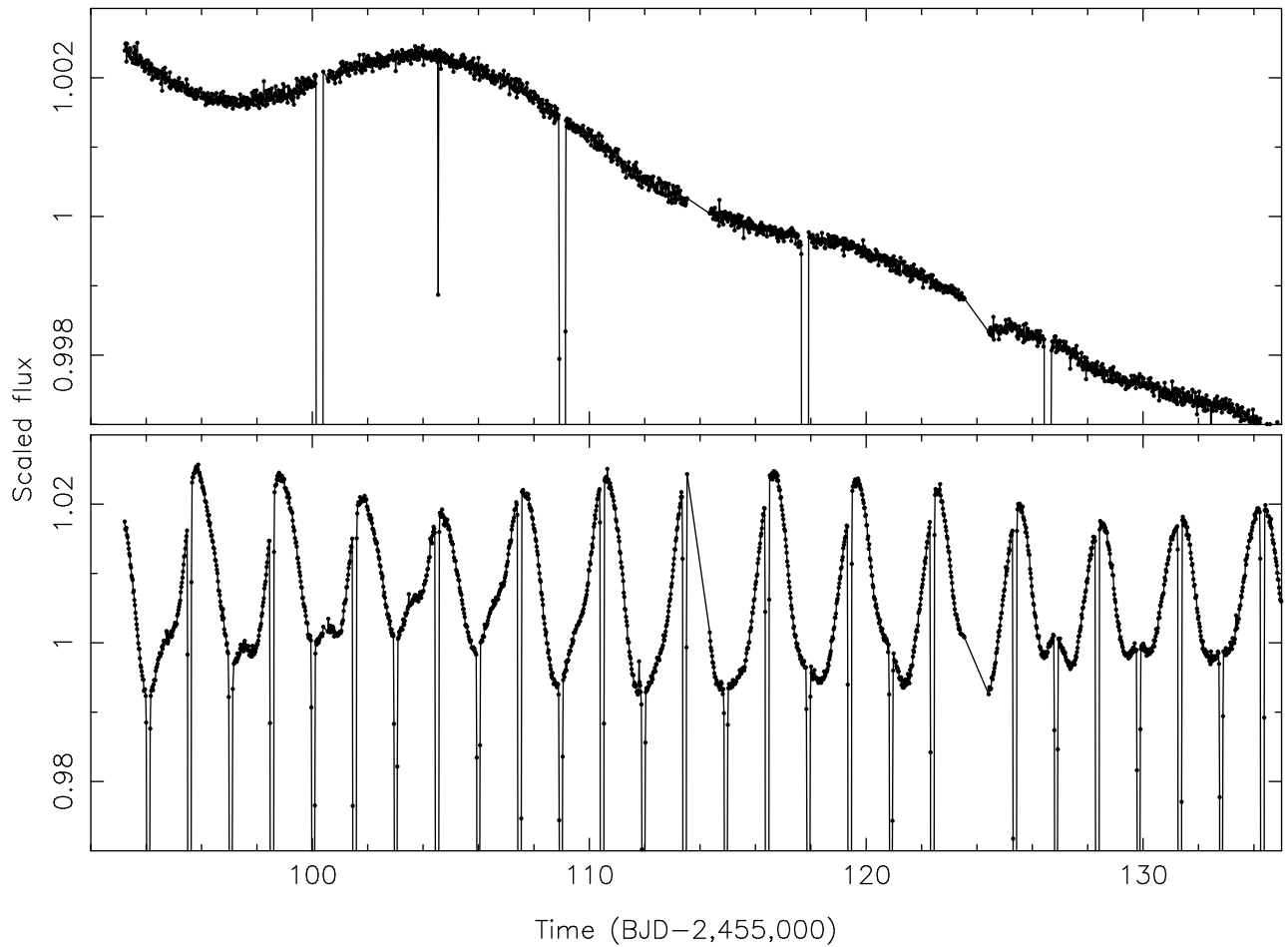}
\caption{Top:  A portion of the scaled
SAP light curve of KIC 6131659
from Q3.  The dip near day 104.5 is a single
point, and is likely an instrumental artifact.
In addition to long-term decrease that
is probably instrumental, there is a
modulation in the out-of-eclipse flux at the level of
0.1 to 0.2\%.
Bottom: A portion of the scaled SAP light curve
of KIC 11228612 from the same time period.  
Note that the $y$-axis scale is
ten times larger compared to the top panel.
Here we see variability in the out-of-eclipse
regions of about 2\%.}
\label{showspot}
\end{figure}

We have found that
the out-of-eclipse variability seen in the {\em Kepler}
light curves
is a good indicator
of star-spot activity.
As seen in Figure \ref{plotraw}, the out-of-eclipse variability
in the SAP
light curve of
KIC 6131659 is at a fairly low level, with a modulation
on a several day time scale
of $\lesssim 0.2\%$ of the mean flux level.  
In Figure \ref{showspot}, we
compare the SAP light curve of KIC 6131659 to that of
the short-period binary KIC 11228612, which is another target in
our sample of objects for ground-based follow-up.  The
period of KIC 1122812 is 2.98046 days, and our spectroscopic results
show that the mass of its secondary is about $0.69\,M_{\odot}$,
which qualifies it as being ``low mass''.
Judging from the light curve, it appears that the rotation of
the star with the spot is roughly synchronized with the orbit, which
is not surprising
given the short orbital period.
The out-of-eclipse variability in KIC 1122812 is about a factor of
10 larger compared to that seen in KIC 6131659. 
For comparison,
\citet{Basri_2011} presented a variability study of a large sample of G and M
dwarfs with {\em Kepler} data.  For stars with roughly periodic
variability, the typical amplitude of variability
seen in the light curves is
$\approx 2$ mmag, which is
similar to the level of of variability seen in KIC 6131659.


\subsection{Ground-based Photometry}

\begin{figure}[h]
\epsscale{0.9}
\includegraphics[angle=-90,scale=0.7]{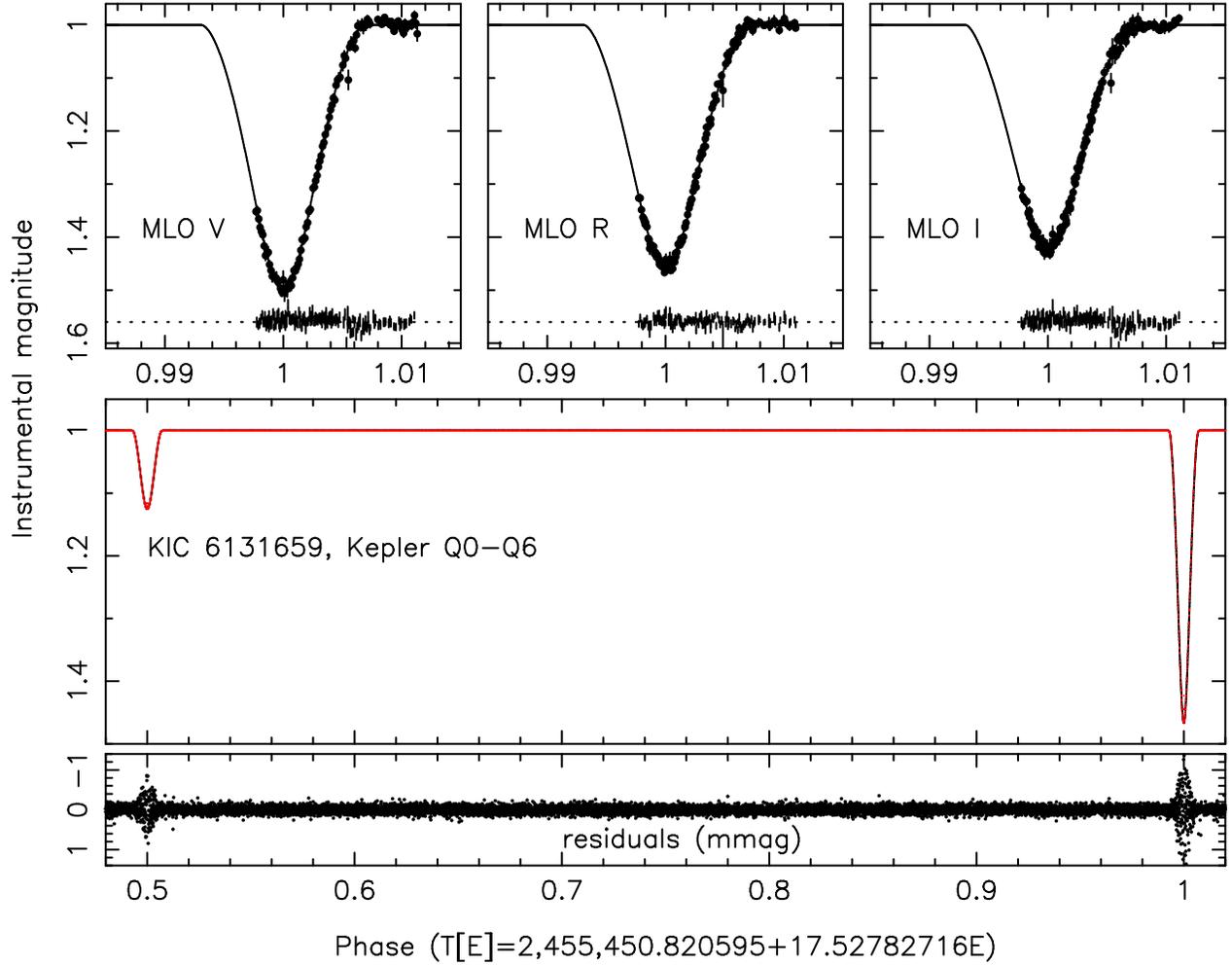}
\caption{Top: the phased Mount Laguna Observatory $V$-,
$R$-, and $I$-band light curves
(taken during primary eclipse) and the best-fitting models.
The residuals of the fits are shown along the dashed lines. 
Middle: The phased, detrended, and normalized
{\em Kepler} light curve (red) and best-fitting model (black). 
Bottom: The residuals of the model fit to the {\em Kepler}
data.}
\label{plotLConly}
\end{figure}

KIC 6131659 was observed on 2010 August 7 
and
2011 July 6 (UT)
using the
0.6m telescope at the Mount Laguna Observatory (near
San Diego, California) and a
SBIG STL-1001 E CCD camera.  Johnson $R$ and  Kron $I$
filters
were used for the 2010 observations, and Johnson $V$, $R$,
and Kron $I$ filters were used for the 2011 observations. 
The exposure times were 60 seconds for $V$, and 30 to 120 seconds
for $R$ and $I$.  A total of 109 useful images in $V$, 164 images
in $R$, and 195 images in $I$ were obtained.
Twilight flats
were taken, and dark exposures to match the exposure times of
the target images were obtained throughout the night.  IRAF was used to
make the standard corrections and calibrations
for the electronic bias, dark current, and flat-field.
The programs
DAOPHOT IIe, ALLSTAR, and DAOMASTER (Stetson 1987,
1992a, 1992b; Stetson, Davis, \& Crabtree 1991) were used
to
obtain the differential
light curves, which are shown at the top of Figure 
\ref{plotLConly}, and presented in Tables \ref{Vdata}, \ref{Rdata}
and \ref{Idata}.

\subsection{Radial and Rotational Velocities}

We obtained 13 echelle spectra 
using
the High Resolution Spectrograph (HRS; \citealt{Tull_1998})
and the Hobby-Eberly Telescope
(HET; \citealt{Ramsey_1998}) 
between 2010 August 18
and 2011 July 10 (UT).
The instrumental configuration consisted of a resolving power
of 30,000, the central echelle rotation angle, the 316 groove
mm$^{-1}$ cross disperser set to give a central wavelength of
6948 \AA, the $2^{\prime\prime}$ science fiber, and two sky fibers.
The exposure times were  600 seconds, split  into two parts
of 300 seconds each to facilitate the removal of cosmic rays.
After the electronic bias was removed from each image, pairs
of images were combined using the ``crreject'' option for
cosmic ray removal.  Spectra were extracted from the resulting
two dimensional images using the ``echelle'' package in IRAF.
In this study we used the spectra from the ``blue'' CCD,
which provides a wavelength coverage between about 5100 and
6900 \AA.
The data are generally of high quality, with signal-to-noise
ratios on the order of 100 per pixel near the peak
of the blaze functions.   

\begin{figure}[h]
\epsscale{0.9}
\includegraphics[angle=-90,scale=0.7]{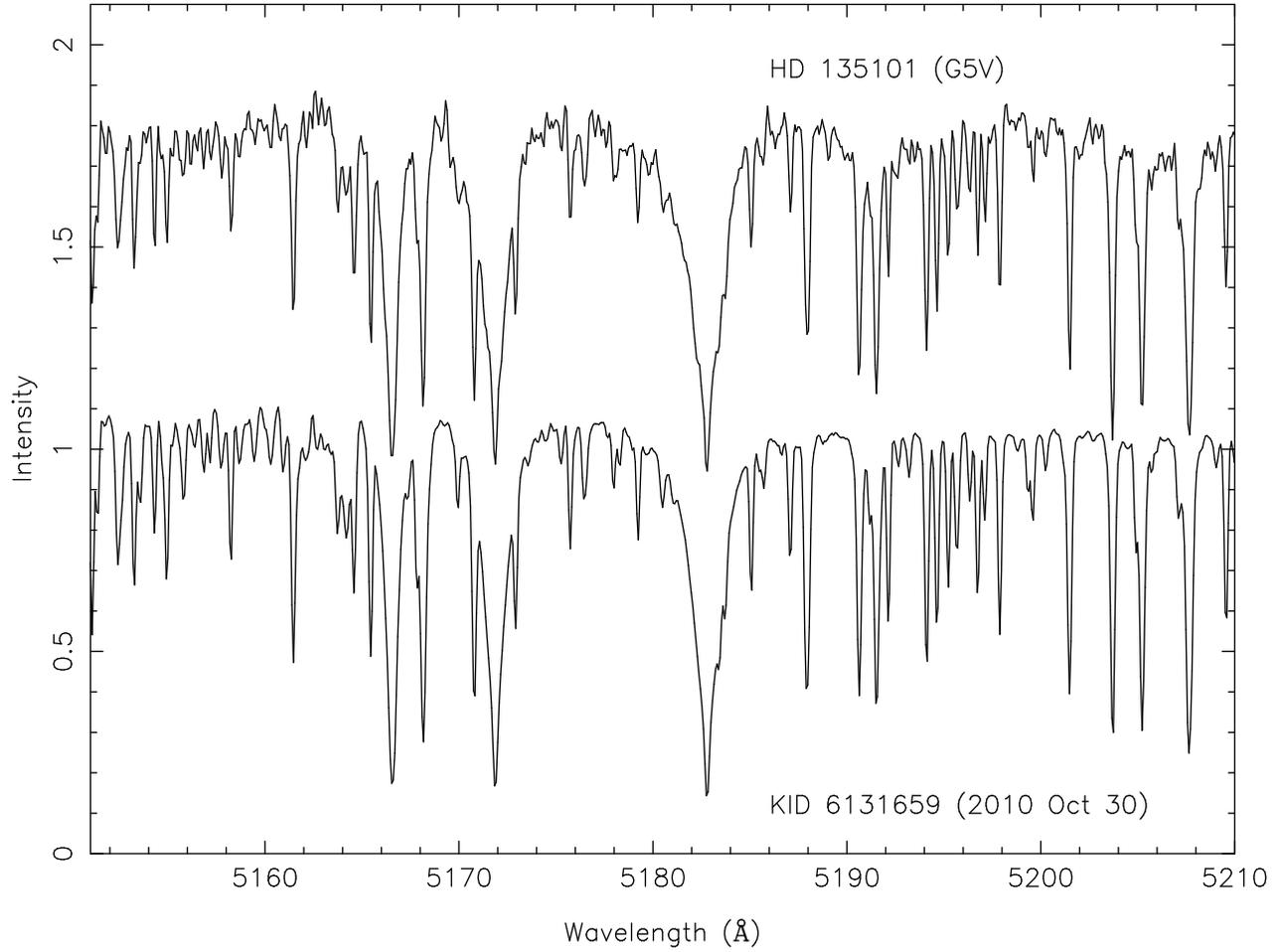}
\caption{The normalized
spectra of the G5V star HD 135101 (top, shifted upward)
and KIC 6131659 (bottom) in the region near the Mg b lines.
This particular observation of KIC 6131659 has been 
Doppler-shifted to match the velocity of HD 135101.}
\label{plotspec}
\end{figure}

Each night, one of
several stars with precise radial velocities taken
from the catalog
of \citet{Nidever_2002} 
were observed by the HET staff using the same instrumental configuration
used for the KIC 6131659 observations.
We have a total of 77 spectra of 40 bright mostly G-type
stars from our program (including ones from
nights when KIC 6131659 was not observed), and using
simple cross correlation near the Mg b
features we found that the spectrum of the G5V star
HD 135101 provided the best match (Figure \ref{plotspec}).

The radial velocities were measured using
the ``broadening function'' technique developed by
\citet{Rucinski_1992,Rucinski_2002}.
The broadening functions (BFs) are  rotational
broadening kernels, where the centroid of the peak yields
the Doppler shift and where the width of the peak is
a measure of the rotational broadening.  For double-lined
spectroscopic binaries the BFs generally provide better
radial velocity measurements than cross-correlation measurements
made with a single template star, especially in cases
where the velocity separation of the two stars is close
to the velocity resolution of the spectra.
See \citet{Bayless_Orosz_2006} for details of this process as
applied to similar HET spectra.
As a check
on the process, we measured the radial velocities
of all of our standard star observations using
the IRAF task fxcor (a cross-correlation routine with
a single template spectrum)
and using the BFs, where the spectrum of
HD 135101 was used as the template in both cases.  
A comparison
of the BF minus fxcor radial velocities shows a median difference
of 0.004 km s$^{-1}$ and an rms of 0.061 
km s$^{-1}$.  The rms of the differences is comparable to the
typical uncertainties in the radial velocity measurements 
reported by fxcor.

\begin{figure}[h]
\epsscale{0.9}
\includegraphics[angle=-90,scale=0.7]{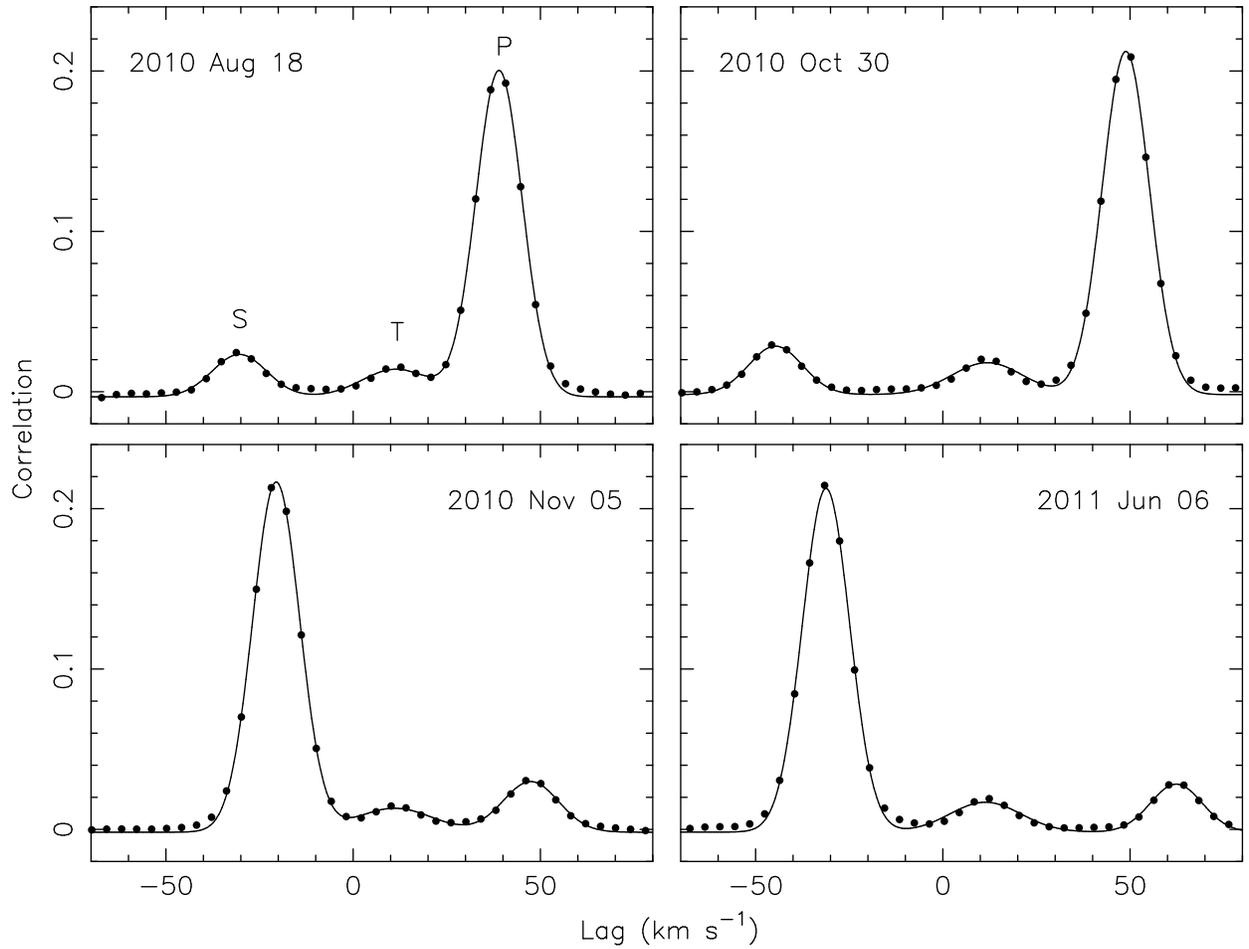}
\caption{The ``broadening functions'' (BFs) from
four observations, with the UT dates indicated.
The lags have been corrected to the heliocentric frame.
The peaks due to the primary (P), secondary (S)
and the third component (T) are indicated in the
upper left panel.}
\label{plotBF}
\end{figure}

We derived BFs for all 13 observations of KIC 6131659 using
HD 135101 
as the template star, whose 
heliocentric radial velocity was taken to be
$-38.921$
km s$^{-1}$  \citep{Nidever_2002}.
We found that
the derived radial velocities are very insensitive to the
spectral type of the template as 
similar
results were obtained using the other
(mainly) G-type radial velocity standard stars and from using
a spectrum of 61 Cyg A (spectral type K5V).  
Figure \ref{plotBF} shows four example BFs.  
Since care
was taken to observe near the quadrature phases, we
see well-defined peaks corresponding to the primary and 
secondary stars.  In every case, there is a third peak located
between the primary and secondary peaks, which occurs at
roughly the same heliocentric velocity.  
For each observation,
we compared the BF derived from the sky-subtracted spectrum to
the one derived from a spectrum with no sky subtraction and found
little, if any change.  Also, the BFs of other targets in
our sample observed with the same instrumentation show
no third peak.  Thus
we believe this third peak is real and is due to a faint
star.

\begin{figure}[h]
\epsscale{0.9}
\includegraphics[angle=-90,scale=0.7]{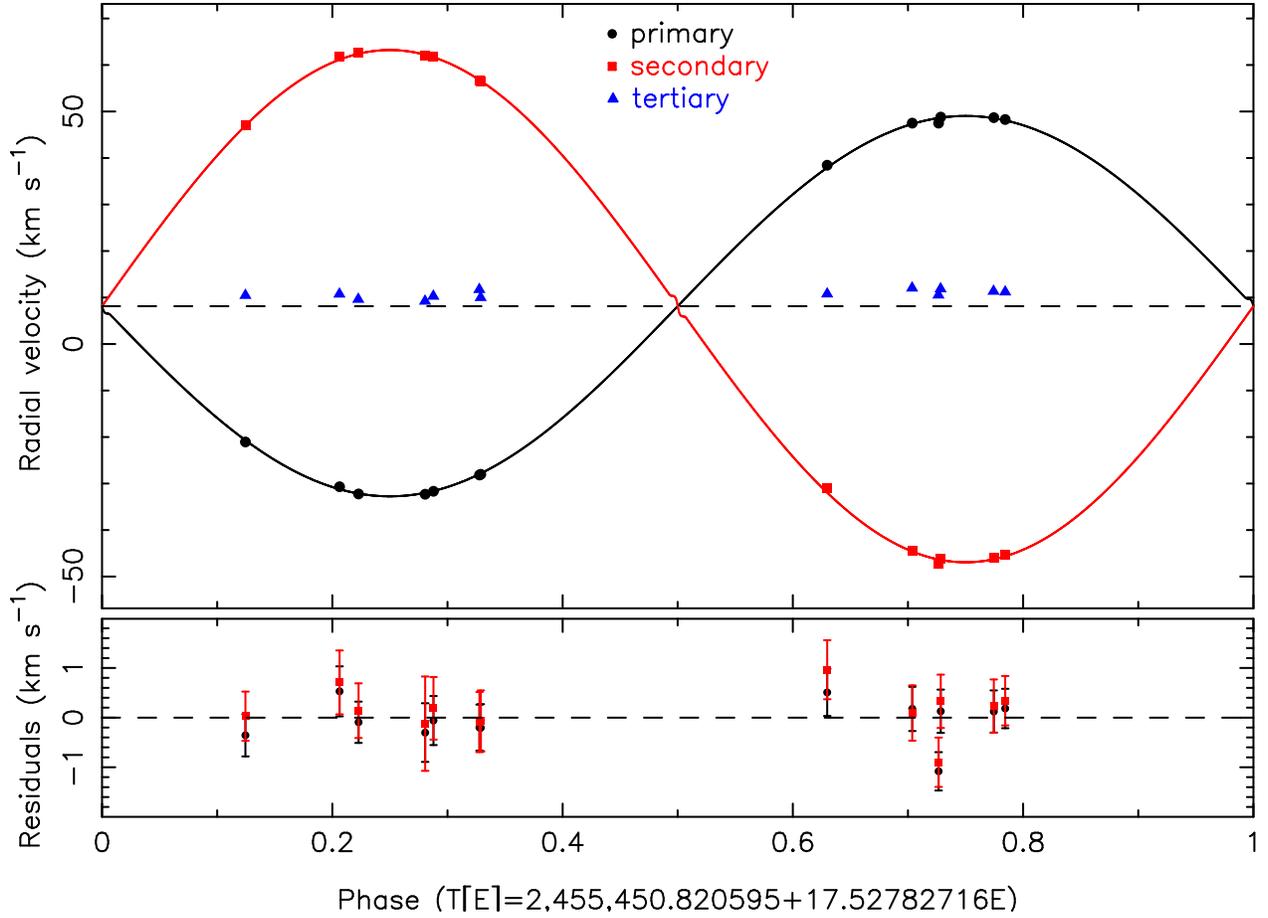}
\caption{Top:  The phased radial velocities
of the primary (filled circles), secondary
(filled squares), and the third component
(filled triangles) and the best-fitting models.
The dashed line indicates the systemic velocity of the system.
Bottom:  The residuals for the fits to the primary and secondary
radial velocities.}
\label{plotRVonly}
\end{figure}

The BFs were fit to a model consisting of three Gaussian
functions\footnote{For cases where the width of the BF peak is
similar to the velocity corresponding to a resolution element,
a Gaussian is often a better model than the standard
analytic broadening kernel (Gray 1992).}
to determine the centroids and widths of the 
peaks.
All of the points in each BF were given equal weight, and formal errors
on the centroids were found by scaling the uncertainties to give 
$\chi^2_{\nu}=1$.
By doing this, the
resulting radial velocities derived from
the peak centroids (Figure \ref{plotRVonly})
have formal uncertainties of typically 0.05 km
s$^{-1}$ for the primary and 0.45 km s$^{-1}$ for the secondary, 
respectively. 
The full width at half maximum (FWHM) of the
primary and secondary peaks in the BFs are approximately 
10 km s$^{-1}$,
which is similar to the spectral resolution.  Thus the
stellar
rotational velocities $V_{\rm rot}\sin i$
are not resolved in our observations
and are less than about 10 km s$^{-1}$.

The BFs of the third star are well separated from the BF peaks of the primary and
secondary star.  Thus the radial velocities of the primary and secondary should
not be biased by the third star. Also, the radial velocities of the third star
are roughly constant and there is no indication that the third star 
interacts with the binary.  Thus, it is sufficient to model the radial
velocities of the primary and secondary with a standard binary star model.
In this case, we modeled 
the radial velocities for the primary and secondary
using a double-sine model where the sines have a common
period 
and systemic velocity and phases 180 degrees apart
(the use of a double sine model is justified since
we show below that the eccentricity is essentially zero).  
The uncertainties
on the individual measurements were scaled to give $\chi^2_{\nu}\approx 1$ for
each curve separately.  The resulting median uncertainties are
0.45 km s$^{-1}$ for the primary and 0.57 km s$^{-1}$ for the secondary,
respectively.  Figure \ref{plotRVonly} shows the velocities
and the best-fitting models.  The final adopted radial velocity
measurements with the scaled
uncertainties are given in Table \ref{RVdata}, and the  
spectroscopic orbital parameters are given in 
Table \ref{specparm}.

The HET's HRS used at resolution 
$R=30,000$ can yield radial velocities with
very small errors on sufficiently bright stars,
provided that ThAr comparison lamp spectra are obtained in
a bracketing fashion, both before {\em and}
after observations of celestial targets, to detect and
correct for ``settling'' of the cross-disperser grating
following configuration changes (S. Mahadevan 2011,
private communication).  Configuration changes
are frequent, since the HET is used in a queue-scheduled mode.
In our program, however, ThAr lamp observations were often obtained
only before {\em or} after observations of a target.

\begin{figure}[h]
\epsscale{0.9}
\includegraphics[scale=0.88]{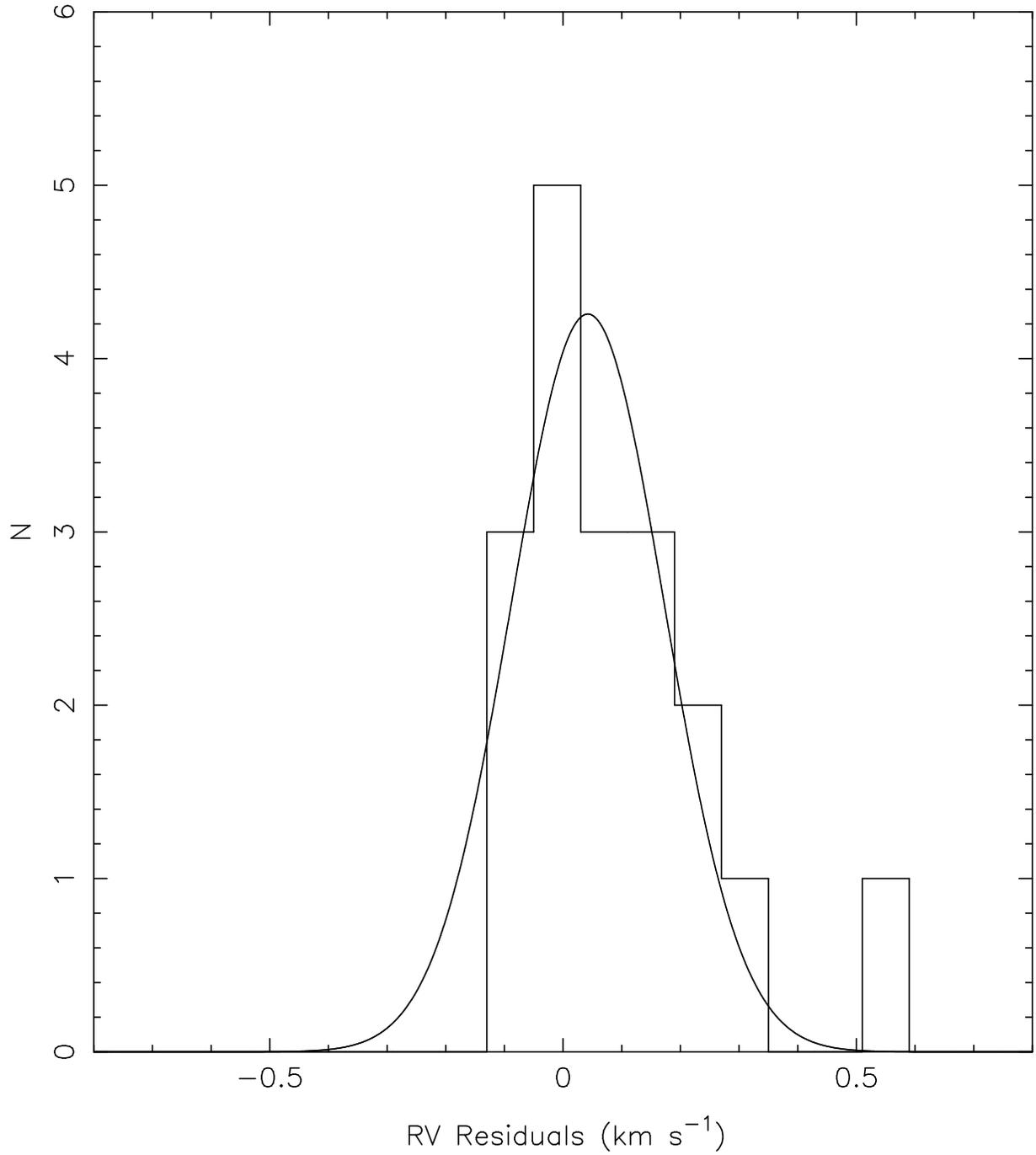}
\caption{The distribution of the 
differences between the measured radial velocity and
the catalog velocity \citep{Nidever_2002}
for the 18 template star observations.  
The median
value is 0.069 km s$^{-1}$ and the rms is
0.157 km s$^{-1}$.
The smooth
curve is a Gaussian with a
$\sigma$ of 0.130 km s$^{-1}$.
}
\label{OCRV}
\end{figure}

As noted before we have 77 observations of 40 bright stars in our
program.
We established a grading scale to quantify the stability of the
HRS cross-disperser during the observation of the standard stars
and their associated ThAr arc exposures, based mainly on the length
of time between the configuration change and the actual
observations.  Nineteen observations were deemed to be
trustworthy in the sense that there probably was enough
time for the cross disperser to settle before the
standard star observation (our best-matched template
star HD 135101 is included in that sample).  In most of these
cases there was an observation of a rapidly rotating B-star
or other type of calibration observation just after
the configuration change and before the standard star observation.
Using HD 135101 as the template, we measured the radial velocity of
the other 18 stars in this sample and compared the measured
velocities with the velocities from the catalog of 
\citet{Nidever_2002}.   Although this catalog does not give uncertainties
on the individual velocities, the rms scatter of multiple measurements
for each star is reported to be less than 100 m s$^{-1}$.
A histogram showing the distribution
of radial velocity differences
is shown in Figure \ref{OCRV}.  
The median difference is 0.069 km s$^{-1}$
and the rms is 0.157 km s$^{-1}$.  
The distribution can be fit by a Gaussian with
a $\sigma$ of 0.130 km s$^{-1}$.  
Based on this analysis we conclude
the zero-point of our
velocity  is consistent with
the scale
defined by the \citet{Nidever_2002} catalog.  Furthermore,
based on the spread of the radial velocity
residual values,  the floor on the 
uncertainties
of individual velocity measurements is
$\approx 0.16$ km s$^{-1}$.

The radial velocities of the third star show no obvious trend.
The median heliocentric velocity of the third star is 10.74 km s$^{-1}$, 
and the rms of the 13 measurements is 0.83 km s$^{-1}$.  This median
velocity is  close to the systemic velocity of the binary,
which is $\gamma=8.142\pm 0.008$ km s$^{-1}$, where the uncertainty
does not include the error in the velocity zero-point.
When one considers the
velocity dispersion
of stars in the solar neighborhood, which is anywhere between $\approx 20$
and 60 km s$^{-1}$,  depending on the age
\citep{Nordstrom_2004}, the chance that two random stars will have a radial 
velocity within $\approx 3$ km s$^{-1}$ of each other is somewhat small.
Also, a wide, hierarchical triple with an outer period of several years could have
a velocity difference of a few km s$^{-1}$, so it is tempting to conclude
the third star is bound to the eclipsing binary.  
However,
the modeling results presented below seem to suggest
that the third star is not related to the binary.

\subsection{Metallicity and Effective Temperatures}

Accurate temperatures and metallicity are essential for the
characterization of both stars, but our photometric data does not
provide strong constraints on either parameter. The eclipses observed
in the light curve yield the {\em ratio} $T_{{\rm
eff},2}/T_{{\rm eff},1}$, but only weakly constrain the absolute
temperatures, and the metallicity cannot be reliably determined
photometrically. A spectroscopic analysis can determine the effective
temperature, surface gravity, and metallicity, but all three
parameters are highly correlated and the results are unreliable in the
absence of external constraints. In eclipsing systems the surface
gravities can be determined from the light curve, effectively reducing
the problem to a more manageable $T_{\rm eff}-{\rm [m/H]}$
degeneracy. Similar to the spectroscopic analysis of Kepler-34 and
Kepler-35 \citep{wel12}, we employed the two dimensional cross-correlation
routine TODCOR \citep{Zucker_1994} and its extension into three dimensions
TRICOR,
along with the Harvard-Smithsonian Center for
Astrophysics (CfA) library of synthetic spectra to determine the
effective temperatures of the binary members and the system
metallicity.

The CfA library consists of a grid of Kurucz model atmospheres
\citep{kur05} calculated by John Laird for a linelist compiled by Jon
Morse. The spectra cover a wavelength range of $5050-5360$ \AA, and
have spacing of $250$~K in $T_{\rm eff}$ and $0.5$ dex in $\log g$ and
[m/H]. Initially, TODCOR was used to get the parameters for the primary
and secondary.  To do this, 
we cross-correlated the HET/HRS spectra with every pair of
templates spanning the range $T_{{\rm eff},1}=[4000,6500]$, $T_{{\rm
eff},2}=[3500,6000]$, $\log g_{1,2}=[3.0,5.0]$, ${\rm
[m/H]}=[-1.5,+0.5]$, and recorded the mean peak correlation
coefficient at each grid point. Next, we interpolated to the peak
correlation value in each parameter (but fixed the surface gravities
to those found from the photometric analysis) to determine the
best-fit parameters for the binary. 
After the initial TODCOR fits, the third star was accounted for
using TRICOR, which yielded improved  parameters for the stars
in the binary.
Given the coarseness of the
template grid, we assigned internal errors of $100$~K in $T_{\rm eff}$
and $0.20$ dex in [m/H]. As mentioned above however, the degeneracy
between temperature and metallicity could cause correlated errors
beyond those quoted here. We explored this by fixing the metallicity
to the extremes of the $1$-$\sigma$ errors and assessing the resulting
temperature offset. Incorporating these correlated errors, we report
the final parameters for 
KIC 6131659: 
$T_{{\rm eff},1} = 5660 \pm 140$~K,
$T_{{\rm eff},2} = 4780 \pm 105$~K, ${\rm [m/H]} = -0.23 \pm 0.20$
dex. The calculated light ratio in the wavelength range used in the
analysis ($5150-5360$ \AA) was $L_2/L_1 = 0.105 \pm 0.004$.
The derived radial velocities were consistent with those
derived using the BF analysis, where $K_{1,{\rm TRICOR}}
=40.95\pm 0.18$ km s$^{-1}$ and $K_{2,{\rm TRICOR}}=
54.90\pm 0.22$ km s$^{-1}$.

The temperature, gravity, and metallicity of the third star
are not well constrained by the TRICOR analysis.  The third
star contributes relatively little to the fits, and similar
correlations were found even when the temperature and
gravity of the third star were changed by several grid points.
In the end, values of $T_3=5000$~K, $\log g_3=3.5$, and
$[m/H]_3=0$ were adopted for the final TRICOR runs.

\section{Light Curve Modeling}
\label{sec:DataAnalysis}

\subsection{Overview}

We use the ELC (Eclipsing Light Curve) code \citep{Orosz_2000} to model
the light curves.  ELC normally uses the standard Roche geometry to specify
the shapes of the stars, and uses specific intensities from model
atmospheres.  The stellar surfaces are divided up into tiles, and
the flux is obtained by numerical integration.  When fitting high
signal-to-noise light curves such as those provided by {\em Kepler},
one obviously needs high precision models, and this is
usually done by using more tiles over the stellar surfaces.  However,
the computational
time needed to fully sample parameter space becomes prohibitive,
so for well-detached binaries 
such as KIC 6131659 where the stars are very nearly perfect spheres, 
another mode of operation was added to ELC where the
analytic expressions developed by \citet{Kopal_1979}, 
and revisited
by \citet{Gimenez_2006}, are used to compute the light curves.  Those equations
are a function of the fractional radii of the two stars, the inclination,
the limb darkening coefficients, and the orbital phase.  For each star,
the specific intensity at the normal (i.e.\ at $\mu=1$) is taken
from the model atmosphere table for the appropriate temperature and gravity,
and intensities for other angles are computed from the limb
darkening law, which is either the linear 
[$I(\mu)=I_0(1-x(1-\mu)$]
or the ``quad'' 
[$I(\mu)=I_0(1-x(1-\mu)-y(1-\mu)^2)$]
law.
ELC's ``analytic'' mode has the advantage in that it is very fast, and
large swaths of parameter space can be explored in a relatively short
time.  The disadvantages are that 
other effects, such as
reflection, ellipsoidal
variations, and star spots
cannot be easily modeled.   

In order to fully match the {\em Kepler} long cadence (LC) data,
the model light curve is computed at phase intervals that
correspond to roughly 1 minute. The model {\em Kepler}
curve is then binned using simple numerical integration
to a bin size of 29.4244 minutes \citep{Gilliland_2010}.  Such binning
is absolutely necessary as the shapes of the model eclipse profiles
change noticeably.  Since the ground-based light curves are measured
using images with much shorter exposure times, no binning is applied
for them.

For KIC 6131659, we used a model with 24 free parameters.
The inclination $i$, mass ratio $Q=M_2/M_1$, the $K$-velocity
of the primary, and the orbital period $P$
set the scale of the binary (e.g.\ the semimajor axis
$a$ is uniquely determined).  The shape and orientation
of the orbit is set by the
eccentricity $e$, the argument of
periastron $\omega$, and a reference epoch, which we take to be the time
of the primary superior 
conjunction $T_{\rm conj}$.  The fractional radii $R_1/a$ and $R_2/a$
set the sizes of the stars.  The stellar temperatures are determined by 
specifying the primary temperature $T_1$ and the temperature ratio
$T_2/T_1$. For KIC 6131659, we have two external
sources of contaminating light.  There
is light from the  star noticed in the spectroscopy (and possibly physically
associated with the binary) which contaminates all
of the light curves.  This third light is parameterized by the
temperature $T_3$ and a scaling factor $S$, which is the ratio
of the projected area of the third star to the projected area of
the primary.  
There is also contaminating
light in the {\em Kepler} light curve from stars on the sky whose
light leaks into the KIC 6131659 aperture.
According to the
Kepler Input Catalog, the fraction of light from other sources is
a few percent.
Since the {\em Kepler} pixels are relatively
large, the stars that contaminate the {\em Kepler} light curve
are not a problem for ground-based observations, so a correction
is applied only to the model {\em Kepler} light curve.
The
parameter $k$ gives the fractional
level of contamination, and the modified flux
$f_{\rm new}$
at each phase is given by
\begin{equation}
f_{\rm new}=f+{kf_{\rm med}\over 1+k}
\end{equation}
where $f$ is the model flux before correction, and  $f_{\rm med}$
is the median flux of the entire model light curve before correction.
Finally, we have two limb darkening coefficients for each star for
each bandpass.  However, since the secondary eclipse is not observed
in the $V$,
$R$ and $I$ light curves, we don't fit for those coefficients.
This leaves us with a total of 10 coefficients:  $x_1(K)$ and $y_1(K)$
for the primary in the {\em Kepler} bandpass, 
$x_2(K)$ and $y_2(K)$ for the secondary
in the {\em Kepler} bandpass, $x_1(V)$ and $y_1(V)$ for the primary in $V$,
$x_1(R)$ and $y_1(R)$
for the primary in $R$, and $x_1(I)$ and $y_1(I)$ for the primary in $I$.

\subsection{Combined Long Cadence Data Fits}

ELC's genetic algorithm
and its Monte Carlo Markov Chain were
both used to optimize the model fits to the LC data.  Generally speaking, the parameters
were allowed to vary over relatively wide ranges.  The fits
are not sensitive to the adopted value of the primary temperature.
Based on the spectral type of around G5V and on the TODCOR
analysis, we adopted a range
of $5460 \le T_1\le 5860$ K.  
After some initial runs it was apparent that there was some difficulty
with the relative weights between the radial velocity curves and the light curves.
Solutions would often converge where the radial velocities were not optimally
fit.
Some experimentation showed
that the optimization worked better
when the radial velocity curves were decoupled from the
light curves.  Thus
we fit only the photometric data, using the usual $\chi^2$ statistic
as the fitness function:
\begin{eqnarray}
\chi^2_{\rm photo} 
&=& \sum_{i=1}^{23123}{(f(\phi_i;\vec{a})-f_{\rm obs}(K))^2\over \sigma_i^2}
\nonumber \\
&+& \sum_{i=1}^{109}{(f(\phi_i;\vec{a})-f_{\rm obs}(V))^2\over \sigma_i^2(V)}
\nonumber \\
&+& \sum_{i=1}^{133}{(f(\phi_i;\vec{a})-f_{\rm obs}(R))^2\over \sigma_i^2(R)}
\nonumber \\
&+&  \sum_{i=1}^{162}{(f(x_i;\vec{a})-f_{\rm obs}(I))^2\over \sigma_i^2(I)}
\end{eqnarray}
where $f(\phi_i;\vec{a})$ is the model flux at a given phase $\phi_i$ for
a vector of parameters $\vec{a}$, $f_{\rm obs}$ is the observed
value at the same phase, and  $\sigma_i$ is the uncertainty.
The uncertainties on the measurements were scaled to give 
$\chi^2\approx N$ for each data set separately.  The resulting median
uncertainties are 0.000089 mag, 0.0076 mag,
0.0070 mag, and 0.0090 mag for the {\em Kepler}, $V$, $R$, and $I$-band
light curves, respectively.
The component masses were forced
to be consistent with the values found from the separate fits to
the radial velocity curves by means of additional
$\chi^2$ penalties:
$$
\chi^2_{\rm RV}=(M_1-0.922)^2/0.007^2+(M_2-0.685)^2/0.005^2.
$$
The fitness function used  for a given model was
$\chi^2_{\rm tot}=\chi_{\rm photo}^2+\chi_{\rm RV}^2$.
Roughly 2.2 million  models were computed from
runs of the genetic and Markov chain codes to arrive at the
optimal model.  The uncertainties on the fitted parameters
and derived astrophysical parameters were arrived at by marginalizing
the $\chi^2$ hypersurfaces over the various parameters of interest.
The best-fitting models are shown in Figure \ref{plotLConly} 
and the parameters
are summarized in Table \ref{lightparm}.
The orbital eccentricity is small, but differs from zero.
Since the argument of periastron $\omega$
and the eccentricity $e$
are
highly correlated, we give the quantity $e\cos\omega$
in Table \ref{lightparm}.
The contamination in the {\em Kepler} light curves due to other
light sources leaking into the aperture
is found to be zero within the errors.  

\subsection{Long Cadence Fits to Individual Eclipses}

\begin{figure}[h]
\epsscale{0.9}
\includegraphics[angle=-90,scale=0.7]{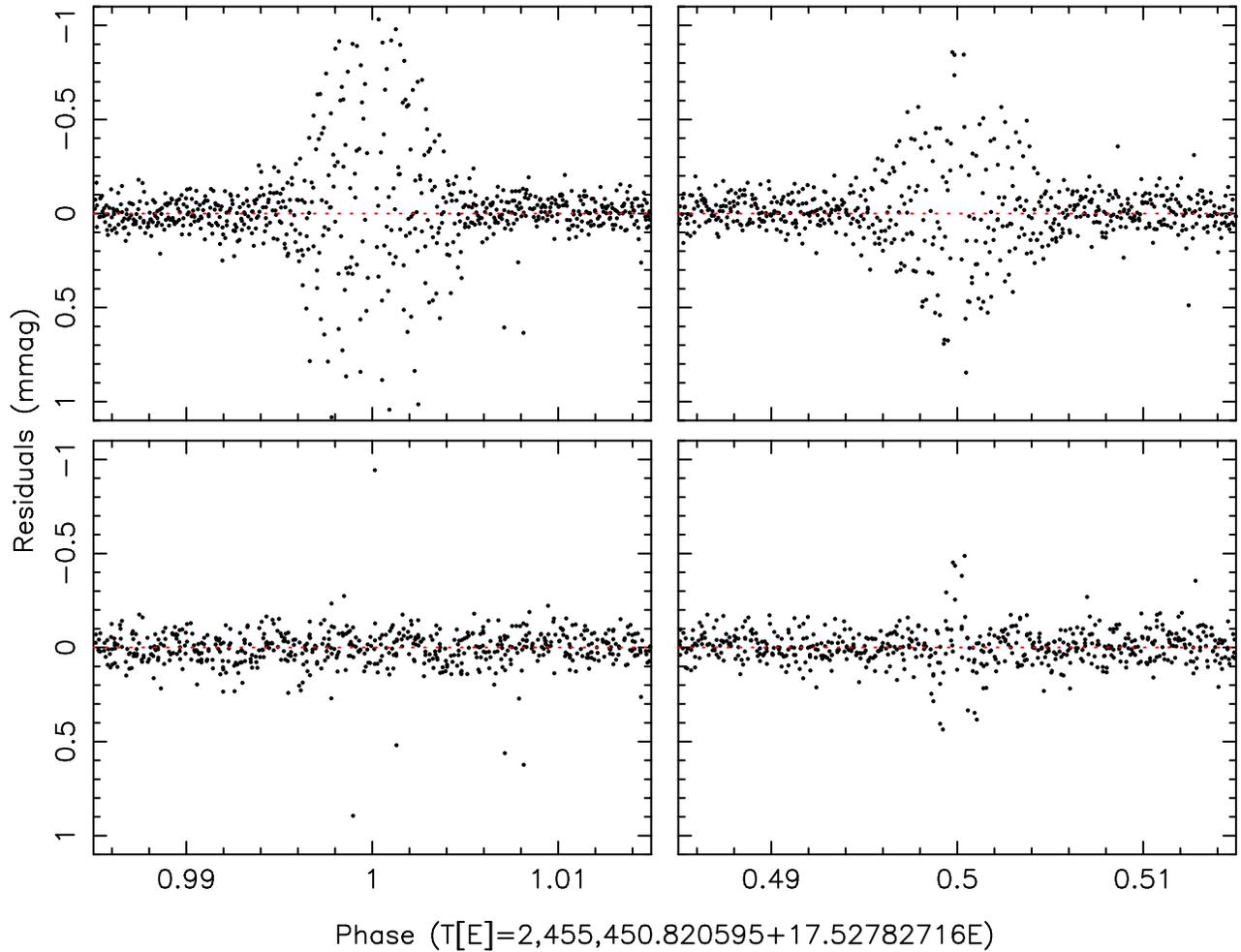}
\caption{Top:  The residuals (in mmag)
of the combined fit to the primary eclipse (left)
and secondary eclipse (right).  The
scatter in the out-of-eclipse regions is considerably
smaller than it is during eclipse, suggesting systematic
issues.    
Bottom:  The residuals (in mmag) obtained after fitting
individual pairs of primary and secondary eclipse
separately.  Here the scatter in the out-of-eclipse phases
matches the scatter during eclipse.}
\label{plotresiduals}
\end{figure}

\begin{figure}[h]
\epsscale{0.9}
\centerline{\includegraphics[scale=0.8]{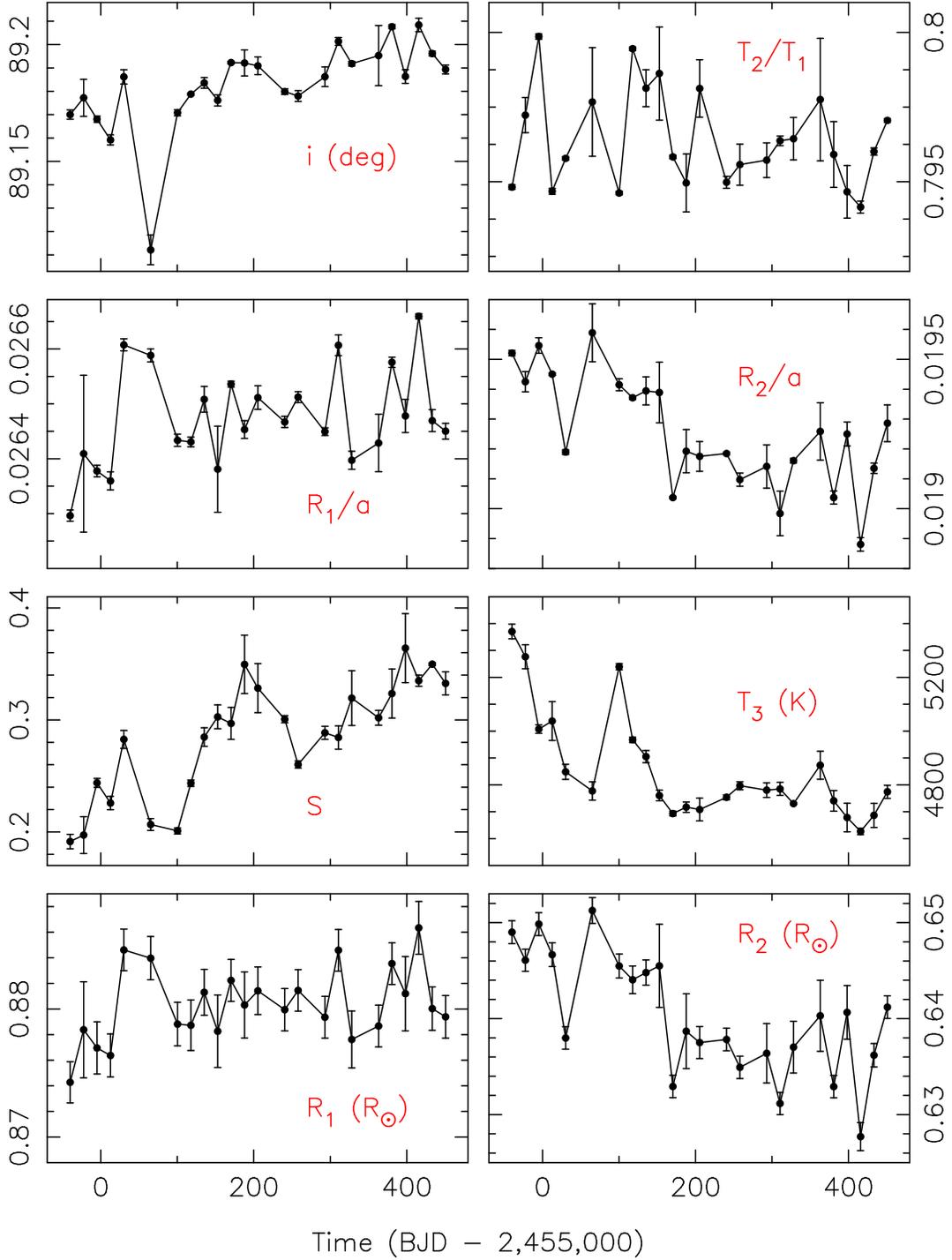}}
\caption{Various fitting and derived parameters derived
from the individual light curve segments are shown as a function of
the primary eclipse time.
From the top, the inclination in degrees, the temperature ratio
$T_2/T_1$, the fractional radii $R_1/a$ and $R_2/a$, the
third light scaling $S$, the third light temperature $T_3$ in K,
and the primary radius $R_1$ in $R_{\odot}$
and the secondary radius
$R_2$ in $R_{\odot}$ are
shown. 
\label{plotspread}}
\end{figure}

Figure \ref{plotresiduals} (top panels)
shows the residuals from the combined
Q0-Q6 fit for both the primary eclipse and secondary eclipse.
The scatter in the out-of-eclipse regions is somewhat smaller
than the scatter in the eclipse phases, which suggests that
there may be some
small systematic errors.  As a check on these systematic issues, 
we divided the
{\em Kepler} light curve into
24 segments, each containing 
a pair of primary and secondary eclipses.    Each segment was
paired with the ground-based $V$, $R$ and $I$ light curves
and modeled separately.  Figure \ref{plotresiduals} 
(bottom panels)
shows
the stacked residuals from all 24 fits.  In this case the scatter
in the in-eclipse and out-of-eclipse phases is similar.
Evidently there are small changes from eclipse to eclipse,
perhaps caused by small star spots being eclipsed, errors
in the light curve normalization, or changes in the contamination level
from Quarter-to-Quarter, or all three.  The change to each
eclipse profile is relatively minor and can be fit for
by small changes in the fitting parameters such as
the inclination or fractional radii.
Table \ref{lightparm} gives the mean,
minimum and maximum of each free parameter over the 24 segments,
as well as the rms.  Figure \ref{plotspread} shows the individual values
plotted as a function of time for eight fitted and derived parameters.
In most cases, the rms of the individual values from the 24 segments
is a bit larger than the formal $1\sigma$ error derived from the combined
{\em Kepler} light curve, which might be an indication that the formal
uncertainties on the values derived from the combined fit are too small.
We note that 
although the fits to each segment are
not completely independent owing to the inclusion of the same
ground-based data in each one, the spread in the resulting
best-fitting parameters that are strongly constrained 
by the {\em Kepler} data, such as the inclination or the fractional
radii, should give us some independent measure of the 
uncertainties.   For our final adopted parameters, in most cases
we adopt the values found from the median of the 24 
individual segment fits, and adopt the rms in those
24 values as conservative $1\sigma$ error.  By doing this,
for the vast majority of the cases,
the difference between the parameter
value derived from the combined Q0-Q6 light curve 
and the median value found from the individual segments
is less than 
$1\sigma$,
using the rms as that uncertainty.

\subsection{Fits to Short Cadence Data}\label{SC}

There is one month of short cadence (SC) data \citep{Gilliland_2010}
for KIC 6131659 publically available, namely from the first month of
Q5.  One primary eclipse and two secondary eclipses are covered.
As noted above, the SC data cannot be modeled simultaneously
with the LC data owing to the vastly different ``exposure times''
(i.e.\ 58.85 seconds vs.\ 29.4244 minutes).  Thus the SC data 
were treated as one of the individual segments discussed above
and modeled with the ground-based light curves.  Table 
\ref{lightparm} gives the parameters.  For the most part there is
very good agreement between parameters measured from the SC data,
the combined LC data, and the individual segments of LC data.  

\begin{figure}[h]
\epsscale{0.9}
\centerline{\includegraphics[scale=0.8]{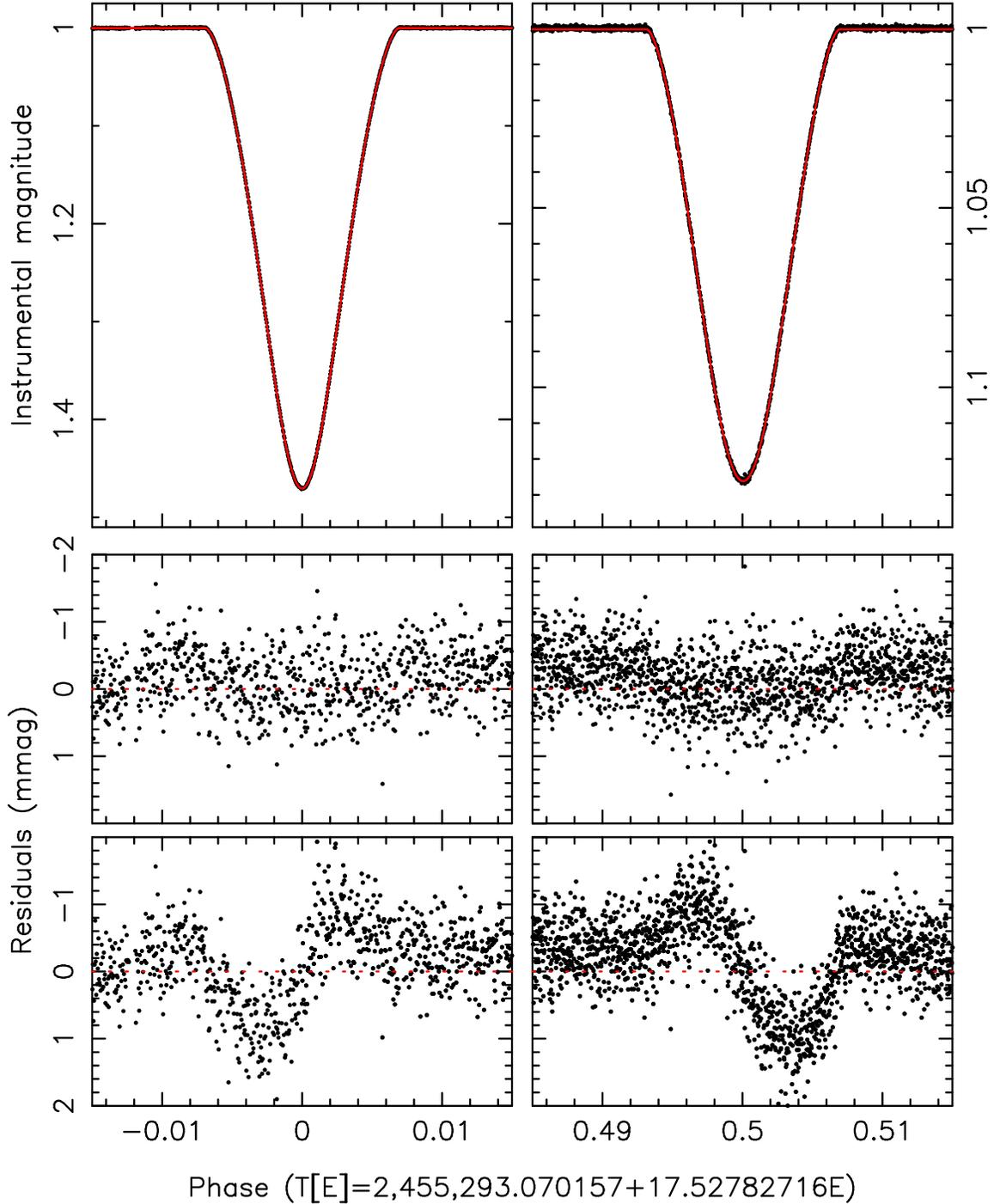}}
\caption{Top:  The folded SC data (black) and
the best-fitting model (red) of the primary eclipse (left) and secondary eclipse (right).
Middle:  The residuals (in mmag) of the fit using
an eccentric orbit.  There are no apparent features
near the eclipse phases.
Bottom:  The residuals (in mmag) obtained after fitting
a model with a circular orbit. In this case there are
distinctive features at the eclipse phases because
the phase difference between the eclipses in the model
does not match the phase difference in the observations.
}
\label{SCresiduals}
\end{figure}

One interesting thing to note about the SC fits is that the
eccentricity is  small,
but is distinctly different from zero:  
$e\cos\omega=(6.88\pm 0.02)\times 10^{-5}$.
Fits with the eccentricity fixed at zero (i.e.\ circular orbit models)
gave significantly worse fits. 
Figure \ref{SCresiduals} illustrates the situation.  
The top panels
show the folded SC data and the best-fitting models 
of both the primary eclipse (left panel) and secondary (right panel).  The
middle panels show the residuals of the fits for the primary
eclipse (left panel) and the secondary eclipse (right panel), and there
is a small scatter with an rms on the order of 1 mmag, and no apparent
features near the eclipse phases.  On the other hand, the residuals
from the circular orbit model have distinctive features near the
eclipse phases (bottom left for the primary eclipse and bottom
right for the secondary eclipse).  The phase difference between the primary
and secondary eclipse for the eccentric model is 0.500044, or about 
67.2 seconds late compared to phase 0.5.  

The delay in the secondary eclipse 
relative to phase 0.5 is due to a combination
of light travel time across the orbit and 
geometry in a slightly eccentric orbit
where the true anomaly is not equal to the mean anomaly.
The time delay due to light travel time is given by
\begin{equation}
\Delta t_{\rm LT}={PK_2\over \pi c}\left(1-{M_2\over M_1}\right)
\end{equation}
where $c$ is the speed of light \citep{Kaplan_2010}.  
Substituting the appropriate
numbers from Table \ref{specparm}, we find $\Delta t_{\rm LT}=23.1$ 
seconds.  Thus
the delay due to light travel time is roughly one third of the 
measured delay.
The change in the relative timing of primary and secondary eclipse is given
by
\begin{equation}
\Delta t_e\approx {2Pe\over\pi}\cos\omega
\end{equation}
\cite{Kaplan_2010}, where a factor of two correction has been applied to his
Equation (6). 
Using values in Table \ref{lightparm}, the above approximation yields
$\Delta t_e=66.4$ 
seconds, which is
very close to the value of 67.2 seconds 
we measured using the short cadence data.  
Accounting for light travel
time, the true delay is 
$\Delta t_{e,{\rm cor}}=44.1$ seconds.  
This then gives
$e\cos\omega=4.57\times10^{-5}$.
Lucy (2012) recently discussed the process of
setting upper limits on the eccentricity where a Bayesian prior
on expected tidal evolution of the eccentricity is included.
Binaries where $e\ll1$ but $e\ne 0$ are occasionally expected
to occur.

\subsection{Discussion of Third Light}

We have two distinct parameterizations of third light.
The {\em Kepler} contamination parameter
$k$ accounts for other unrelated stars that are inside the {\em Kepler}
photometric aperture, but {\em not} inside any ground-based apertures.
As we discussed earlier, $k$ is very close to zero.  
The quantities $S$ and $T_3$ are used to characterize light from a third
star that is present in {\em all} 
of the light curves.  There are strong and compelling
reasons for including such a star, chief among them the fact that a third star
was noticed in the spectra.  

Regarding the parameters for that third star noticed in the spectra,
the scaling parameter
$S$ and the third light temperature $T_3$ are somewhat anti-correlated.  However,
the combined effect of these parameters is
such that the amount of light from the third star is roughly equal
to the amount of light from the secondary star.
This rough equality between the two 
is consistent
with what is seen in the broadening functions
(Figure \ref{plotBF}).   
The value of the parameter $S$ indicates the third star's
{\em angular} radius is between $\approx 44\%$---$55\%$ the 
{\em angular} radius of
the primary.  For comparison, the angular (and physical) radius of
the secondary is $\approx 72\%$ of the primary's radius.
In addition, the temperature of the third star is somewhat
hotter
than the secondary (e.g.\ $T_3=4700-5400$ K vs.\ $T_2=4600$ K).
If the third star was bound to the eclipsing binary, its 
physical radius would be only $\approx 50/72=69\%$ the radius of
the secondary star.  Given that small size, it does not
seem likely the third star would be able to produce the same
{\em flux} as the secondary star if both stars are on the main
sequence.  On the other hand, if the third star were more
distant than the binary, then it could be more luminous than the
secondary (as its hotter temperature suggests it should be) but
at the same time have roughly the same flux. Thus
the fact that
the third star has a radial velocity similar to that of the
binary may be a coincidence.  

Since ELC uses model-atmosphere
specific intensities, the overall third light level should be
reasonably well constrained given the wide bandpass coverage
of the available light curves.  Also, wide ranges for
the third light parameters were searched, and the solutions
converged on a situation where the third light contamination is
far from zero.  
Thus we believe the radii we derive
for the primary and secondary should not be biased.

\section{Comparison With Model Isochrones and Discussion}\label{results}

\begin{figure}[h]
\includegraphics[scale=0.6]{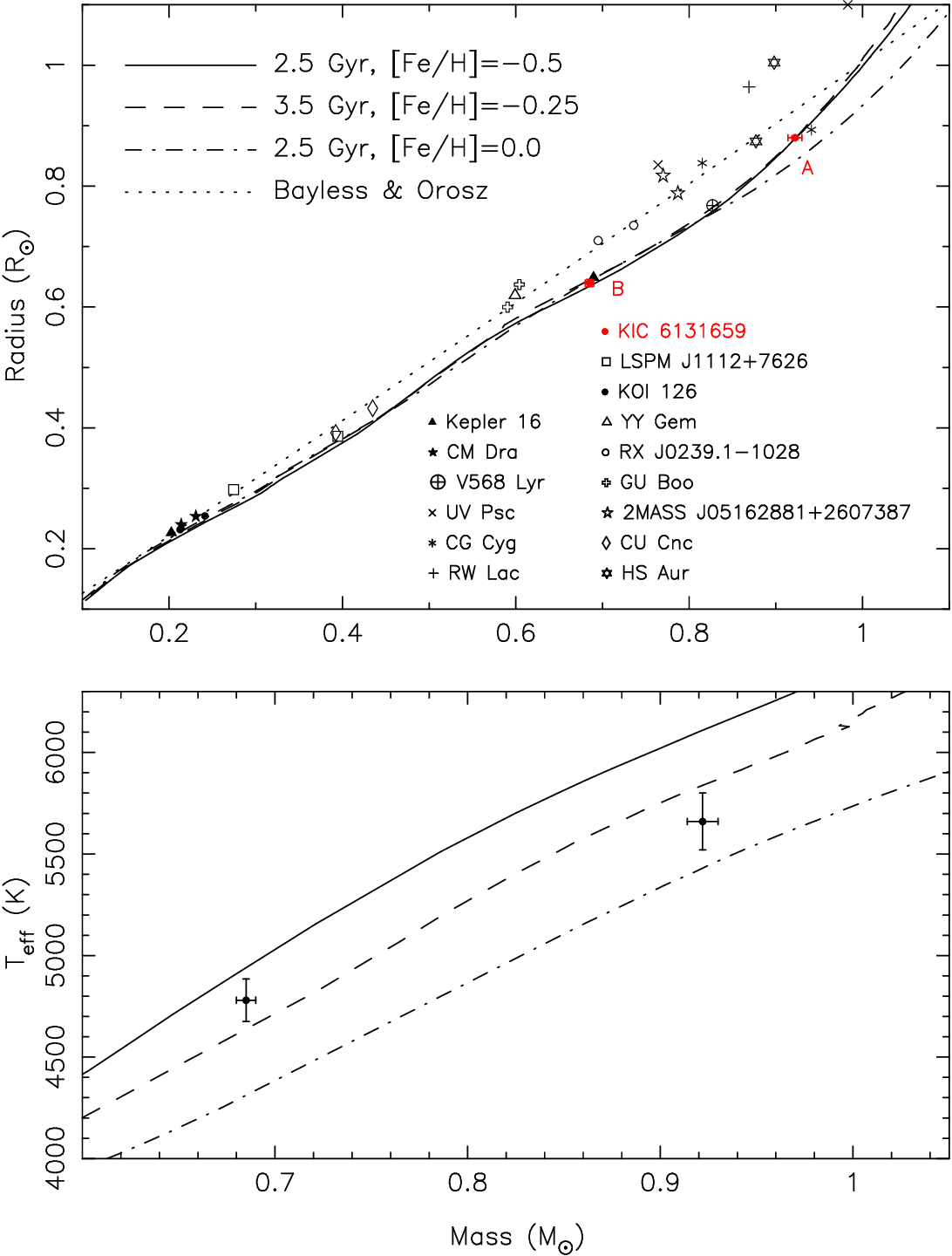}
\caption{Top: A mass-radius diagram
for well-characterized low-mass stars in
eclipsing binaries (measurements from \citet{Torres_2010} and
cited references, \citet{Carter_2011}, and \citet{Irwin_2011}).  
The evolutionary models are from 
\citet{Dotter_2008}.  
Also shown is the empirical relation
derived by \citet{Bayless_Orosz_2006}.
The location of the components in KIC 6131659 are shown in red with the primary denoted by an 'A' and the secondary denoted by a 'B'. Note that the symbol for KIC 6131659 B nearly covers up the symbol for the primary in Kepler-16. The mass-radius relationship from other studies are included on the figure for comparsion. Only well-studied figures with low error bars are included. Some recent work has identified more potential low mass binaries for study \citep[see e.g.][]{Bhatti_2010,Becker_2011}, but have not yet produced well constrained values for the masses for these targets, and are thus not included.
Bottom:  A mass-temperature diagram showing the positions of
KIC 6161659 against the same three isochrones that are displayed
in the top panel, where the effective temperatures were derived
from the TODCOR analysis. 
\label{fig:Isochrone}}
\end{figure}

Figure \ref{fig:Isochrone} shows the positions of the component stars
in KIC 6131659 in a
mass-radius diagram.  The figure also shows three theoretical isochrones
from the Dartmouth Stellar Evolution Program \citep[DSEP,][]{Dotter_2008}
%
and the empirical relationship derived by
\citet{Bayless_Orosz_2006}.  
As noted earlier, most of the low-mass stars
with well-measured masses and radii are above the isochrones,
meaning the stars are larger than expected, given their masses.  
In contrast, both stars in KIC 6131659 are on the same isochrone,
where the one with an age of 3.5 Gyr, [Fe/H]$=-0.25$, 
[$\alpha$/Fe]$=0$, and a helium fraction of
$Y=0.2537$ formally provides the best fit with $\chi^2=0.5287$.
The isochrone with an
age of
2.5 Gyr, [Fe/H]$=-0.5$, 
[$\alpha$/Fe]$=0$, and a helium fraction of
$Y=0.2537$
has a nearly identical $\chi^2$ (0.5397).   The metallicities of the best-fitting
isochrones match very well the metallicity found from the TODCOR analysis
(Table \ref{specparm}).   Figure \ref{fig:Isochrone} also shows a mass-temperature
diagram.  Both the primary and secondary are reasonably close to the best-fitting
isochrone.

\citet{Torres_2007}
introduced a parameter $\beta$ which is the correction factor needed
to make the model predictions of the radius agree with the
measured radius.  These are $\beta_1=1.001\pm 0.003$ for the primary
and $\beta_2=0.992\pm 0.009$ for the secondary, respectively.
These are consistent with unity, meaning no correction is needed to make
the evolutionary models agree with the observations.

Note that
the secondary in KIC
6131659 and the primary in Kepler-16 have 
roughly the same mass as the secondary in
RX J0239.1-1028, whereas the radii differ by $\approx 0.1\,R_{\odot}$
(the secondary in RX J2039.1-1028 is the bloated star).
At these low masses, neither age 
differences nor changes in the metallicity will
account for the large radius discrepancy.
There is therefore at
least one other parameter which influences where a star resides in this
diagram.

With the recent discoveries of the long-period eclipsing binaries
LSPM J1112+7626 \citep{Irwin_2011}, Kepler-16 \citep{Doyle_2011}, and now
KIC 6131659, the observational picture regarding the radii of low-mass
stars is becoming more complex.  The stars in LSPM J1112+7626
are inflated by about 4\%, which indicates that a very
long orbital period by itself does not automatically result in agreement with
the evolutionary models.  \citet{Irwin_2011} also showed that
LSPM J1112+7626 has a 65 day out-of-eclipse modulation of about 2\% which
was attributed to star spot activity.  As we discussed earlier, KIC 6131659
has an out-of-eclipse modulation of $\lesssim0.2$\%, 
which presumably translates to
a lower level of star spot activity.  The primary star in Kepler-16 fits
on the model isochrone in the mass-radius diagram, whereas the secondary
star does not.
There is an $\approx 1\%$ 
peak-to-peak modulation in the out-of-eclipse light curve of Kepler-16
with periodicity of about 35 days
\citep{Winn_2011}, where the primary star is presumed to be the source of the
modulation.  
Given these
three long-period binaries, 
one might further speculate that
there might be a threshold of stellar activity level past which
these stars become inflated.  

\citet{Feiden_2011} have recently presented improved evolutionary models
which were applied to the three stars in the eclipsing triple system
KOI-126 \citep{Carter_2011}.  A model isochrone with
an age of 4.1 Gyr and a metallicity of
$[{\rm Fe/H}]=+0.15$ reproduced the observed radii
of the stars reasonably well, with formal relative errors between the model
predictions 
and observations of $\le 0.3\%$.  
\citet{Feiden_2011} attributed the success of their models to an improved
treatment of the equation of state.  On the other hand, those same
models cannot match the observed radii components
of the well-known low-mass
binary CM Draconis, even when generous uncertainties in the metallicity
are considered.  Although the discovery of Kepler-16 was not announced
when the paper by \citet{Feiden_2011} was accepted for publication, it can be seen
from their Figure 2 that the secondary in Kepler-16 would lie
slightly above the model isochrones in the mass-radius diagram.
This seems to indicate a real dispersion in the observed
mass-radius relation for these fully convective stars.  

\section{Summary and Thoughts on Future Work}\label{sec:Conclusion}

We have analyzed six Quarters of {\em Kepler}
data along with supporting ground-based photometric and
spectroscopic data and obtained precise and
accurate values for the masses and radii for
both stars in the long-period eclipsing binary
KIC 6131659.  The component stars in this binary
are not bloated, in contrast to nearly all other systems.
This suggests that at least one  parameter
in addition to the mass, age, and metallicity strongly
influences the ultimate radius of the star.  

More observational work will be needed to better understand
the mass-radius relation for low-mass stars.  The radii of the
components in
KIC 6131659 agree with the model predictions for a 3.5 Gyr isochrone.
We have shown
that the present level of stellar activity in KIC 6131659 is
relatively low, which seems to suggest that high levels of spot
activity may contribute to the inflated radii found in
the short-period low-mass binaries.  On the other hand,
Kepler-16 has a somewhat higher level of stellar activity, and its
primary star also does not appear to be inflated.  We speculate that perhaps
there is a threshold level of activity past which the stars start to
become inflated.

Although the amplitude of any out-of-eclipse modulations
may not be a perfect indicator
of stellar activity, it is readily observable, especially in 
{\em Kepler } data.  
Are there binaries with very low
levels of out-of-eclipse modulations that nevertheless have stars with
inflated radii?  Similarly, are there additional binaries like
Kepler-16 with spot activity that have at least one star that
is not inflated?    Ground-based follow-up observations of these
binaries, especially the long period systems, can be challenging, 
however, a sample of a dozen systems could lead to a 
clearer understanding
of the mass-radius relation for low-mass stars.


\acknowledgments
We are grateful for the support from the National Science Foundation
via grants AST-0808145, AST-0850564, and AST-0908642,
from the {\em Kepler} Participating Scientist
Program via NASA grant NNX 08AR14G and the from
the {\em Kepler} Guest
Observer program via NASA grant NNX 11AC73G.
Kepler was selected as the 10th mission of the Discovery Program. Funding
for this mission is provided by NASA, Science Mission Directorate. 
The Hobby-Eberly Telescope (HET) is a joint project of the University
of Texas at Austin, the Pennsylvania State University, Stanford
University, Ludwig-Maximilians-Universit\"at M\"unchen, and
Georg-August-Universit\"at G\"ottingen. The HET is named in honor of its
principal benefactors, William P. Hobby and Robert E. Eberly.

\clearpage

\begin{deluxetable}{cc|cc|cc}
\tablecaption{KIC 6131659 $V$-band MLO Photometry\label{Vdata}}
\tabletypesize{\scriptsize}
\tablewidth{0pt}
\tablehead{
\colhead{Time} &
\colhead{Instrumental $V$} &
\colhead{Time} &
\colhead{Instrumental $V$} &
\colhead{Time} &
\colhead{Instrumental $V$} \\
\colhead{(HJD-2,455,000)} &
\colhead{(mag)}  &
\colhead{(HJD-2,455,000)} &
\colhead{(mag)}  &
\colhead{(HJD-2,455,000)} &
\colhead{(mag)}  
}
\startdata
748.75323 & $ 1.3517\pm 0.0076$  &748.82965 & $ 1.3727\pm 0.0091$  &748.91217 &  $ 1.0057\pm 0.0106$  \\ 
748.75525 & $ 1.3497\pm 0.0076$  &748.83167 & $ 1.3517\pm 0.0091$  &748.91418 &  $ 0.9987\pm 0.0076$  \\ 
748.75726 & $ 1.3657\pm 0.0076$  &748.83368 & $ 1.3477\pm 0.0076$  &748.91620 &  $ 1.0007\pm 0.0091$  \\ 
748.75928 & $ 1.3817\pm 0.0076$  &748.83771 & $ 1.3077\pm 0.0091$  &748.91821 &  $ 0.9897\pm 0.0076$  \\ 
748.76129 & $ 1.3897\pm 0.0076$  &748.83972 & $ 1.3057\pm 0.0076$  &748.92023 &  $ 0.9927\pm 0.0091$  \\ 
748.76331 & $ 1.3957\pm 0.0091$  &748.84174 & $ 1.2937\pm 0.0076$  &748.92224 &  $ 0.9977\pm 0.0076$  \\ 
748.76532 & $ 1.4167\pm 0.0091$  &748.84375 & $ 1.2837\pm 0.0076$  &748.93372 &  $ 0.9897\pm 0.0060$  \\ 
748.76733 & $ 1.4347\pm 0.0091$  &748.84576 & $ 1.2677\pm 0.0091$  &748.93573 &  $ 0.9917\pm 0.0060$  \\ 
748.76935 & $ 1.4317\pm 0.0091$  &748.84778 & $ 1.2567\pm 0.0106$  &748.93976 &  $ 0.9987\pm 0.0045$  \\ 
748.77136 & $ 1.4287\pm 0.0106$  &748.84979 & $ 1.2467\pm 0.0091$  &748.94177 &  $ 0.9957\pm 0.0030$  \\ 
748.77338 & $ 1.4517\pm 0.0076$  &748.85181 & $ 1.2287\pm 0.0091$  &748.94379 &  $ 0.9857\pm 0.0045$  \\ 
748.77539 & $ 1.4627\pm 0.0091$  &748.85382 & $ 1.2217\pm 0.0106$  &748.94580 &  $ 0.9937\pm 0.0060$  \\ 
748.77734 & $ 1.4737\pm 0.0106$  &748.85583 & $ 1.2067\pm 0.0091$  &748.94781 &  $ 0.9937\pm 0.0060$  \\ 
748.77936 & $ 1.4727\pm 0.0106$  &748.85986 & $ 1.1927\pm 0.0106$  &748.94983 &  $ 0.9937\pm 0.0045$  \\ 
748.78137 & $ 1.4757\pm 0.0091$  &748.86188 & $ 1.1737\pm 0.0091$  &748.95184 &  $ 0.9997\pm 0.0045$  \\ 
748.78339 & $ 1.4777\pm 0.0121$  &748.86389 & $ 1.1597\pm 0.0106$  &748.95386 &  $ 0.9937\pm 0.0060$  \\ 
748.78540 & $ 1.4837\pm 0.0106$  &748.86591 & $ 1.1507\pm 0.0091$  &748.95587 &  $ 1.0037\pm 0.0045$  \\ 
748.78943 & $ 1.4967\pm 0.0136$  &748.86792 & $ 1.1377\pm 0.0076$  &748.95789 &  $ 1.0117\pm 0.0060$  \\ 
748.79144 & $ 1.5027\pm 0.0091$  &748.86993 & $ 1.1417\pm 0.0076$  &748.95990 &  $ 0.9997\pm 0.0060$  \\ 
748.79346 & $ 1.4807\pm 0.0166$  &748.87195 & $ 1.1137\pm 0.0060$  &748.96191 &  $ 1.0027\pm 0.0045$  \\ 
748.79547 & $ 1.5057\pm 0.0151$  &748.87396 & $ 1.1037\pm 0.0076$  &748.96393 &  $ 0.9927\pm 0.0045$  \\ 
748.79749 & $ 1.4967\pm 0.0151$  &748.87598 & $ 1.1007\pm 0.0106$  &748.96594 &  $ 0.9997\pm 0.0045$  \\ 
748.79950 & $ 1.4977\pm 0.0151$  &748.87799 & $ 1.0987\pm 0.0076$  &748.96790 &  $ 0.9987\pm 0.0045$  \\ 
748.80151 & $ 1.4907\pm 0.0121$  &748.88202 & $ 1.0757\pm 0.0106$  &748.96991 &  $ 1.0097\pm 0.0060$  \\ 
748.80353 & $ 1.4887\pm 0.0136$  &748.88403 & $ 1.0577\pm 0.0091$  &748.97192 &  $ 1.0157\pm 0.0076$  \\ 
748.80554 & $ 1.4917\pm 0.0091$  &748.88605 & $ 1.0627\pm 0.0091$  &748.97394 &  $ 0.9967\pm 0.0076$  \\ 
748.80756 & $ 1.4767\pm 0.0106$  &748.89008 & $ 1.1037\pm 0.0181$  &748.97595 &  $ 1.0047\pm 0.0060$  \\ 
748.80957 & $ 1.4627\pm 0.0076$  &748.89203 & $ 1.0417\pm 0.0060$  &748.97797 &  $ 1.0027\pm 0.0091$  \\ 
748.81158 & $ 1.4637\pm 0.0060$  &748.89404 & $ 1.0347\pm 0.0121$  &748.97998 &  $ 0.9987\pm 0.0060$  \\ 
748.81360 & $ 1.4577\pm 0.0091$  &748.89606 & $ 1.0387\pm 0.0060$  &748.98199 &  $ 1.0007\pm 0.0060$  \\ 
748.81561 & $ 1.4507\pm 0.0091$  &748.89807 & $ 1.0297\pm 0.0076$  &748.98401 &  $ 1.0007\pm 0.0076$  \\ 
748.81763 & $ 1.4417\pm 0.0076$  &748.90009 & $ 1.0437\pm 0.0076$  &748.98602 &  $ 0.9947\pm 0.0091$  \\ 
748.81964 & $ 1.4247\pm 0.0091$  &748.90210 & $ 1.0187\pm 0.0076$  &748.98804 &  $ 0.9817\pm 0.0076$  \\ 
748.82166 & $ 1.4047\pm 0.0076$  &748.90411 & $ 1.0017\pm 0.0106$  &748.99005 &  $ 0.9967\pm 0.0091$  \\ 
748.82367 & $ 1.4047\pm 0.0091$  &748.90613 & $ 1.0007\pm 0.0060$  &748.99207 &  $ 1.0167\pm 0.0136$  \\ 
748.82568 & $ 1.4017\pm 0.0091$  &748.90814 & $ 1.0037\pm 0.0076$  &          &                      \\ 
748.82770 & $ 1.3807\pm 0.0076$  &748.91016 & $ 0.9977\pm 0.0091$  &          &                      \\ 
\enddata
\end{deluxetable}

\clearpage

\begin{deluxetable}{cc|cc|cc}
\tablecaption{KIC 6131659 $R$-band MLO Photometry\label{Rdata}}
\tabletypesize{\scriptsize}
\tablewidth{0pt}
\tablehead{
\colhead{Time} &
\colhead{Instrumental $R$} &
\colhead{Time} &
\colhead{Instrumental $R$} &
\colhead{Time} &
\colhead{Instrumental $R$} \\
\colhead{(HJD-2,455,000)} &
\colhead{(mag)}  &
\colhead{(HJD-2,455,000)} &
\colhead{(mag)}  &
\colhead{(HJD-2,455,000)} &
\colhead{(mag)}  
}
\startdata
415.73306 & $ 1.3659\pm 0.0082$  &415.92581 & $ 0.9920\pm 0.0093$  &748.83038 &  $ 1.3366\pm 0.0058$  \\ 
415.73456 & $ 1.3759\pm 0.0047$  &415.92731 & $ 0.9970\pm 0.0082$  &748.83240 &  $ 1.3286\pm 0.0058$  \\ 
415.74063 & $ 1.4050\pm 0.0047$  &415.93338 & $ 1.0100\pm 0.0070$  &748.83441 &  $ 1.3136\pm 0.0058$  \\ 
415.74216 & $ 1.4130\pm 0.0058$  &415.93488 & $ 0.9970\pm 0.0058$  &748.83643 &  $ 1.3046\pm 0.0070$  \\ 
415.74823 & $ 1.4310\pm 0.0058$  &415.94098 & $ 0.9880\pm 0.0082$  &748.83844 &  $ 1.2846\pm 0.0070$  \\ 
415.74973 & $ 1.4310\pm 0.0047$  &415.94247 & $ 0.9950\pm 0.0082$  &748.84039 &  $ 1.2836\pm 0.0058$  \\ 
415.75580 & $ 1.4510\pm 0.0035$  &415.94858 & $ 1.0029\pm 0.0070$  &748.84241 &  $ 1.2746\pm 0.0082$  \\ 
415.75729 & $ 1.4530\pm 0.0047$  &415.95007 & $ 1.0000\pm 0.0058$  &748.84442 &  $ 1.2526\pm 0.0070$  \\ 
415.76337 & $ 1.4670\pm 0.0047$  &415.95615 & $ 0.9960\pm 0.0058$  &748.84644 &  $ 1.2396\pm 0.0093$  \\ 
415.76489 & $ 1.4629\pm 0.0047$  &415.95764 & $ 1.0059\pm 0.0047$  &748.84845 &  $ 1.2436\pm 0.0070$  \\ 
415.77097 & $ 1.4579\pm 0.0047$  &748.75397 & $ 1.3266\pm 0.0058$  &748.85046 &  $ 1.2296\pm 0.0093$  \\ 
415.77249 & $ 1.4629\pm 0.0047$  &748.75598 & $ 1.3266\pm 0.0070$  &748.85248 &  $ 1.2146\pm 0.0070$  \\ 
415.77856 & $ 1.4469\pm 0.0058$  &748.75793 & $ 1.3486\pm 0.0070$  &748.85449 &  $ 1.2056\pm 0.0058$  \\ 
415.78006 & $ 1.4369\pm 0.0047$  &748.75995 & $ 1.3626\pm 0.0070$  &748.85651 &  $ 1.1926\pm 0.0070$  \\ 
415.78613 & $ 1.4150\pm 0.0058$  &748.76196 & $ 1.3696\pm 0.0070$  &748.85852 &  $ 1.1786\pm 0.0070$  \\ 
415.78763 & $ 1.4030\pm 0.0058$  &748.76398 & $ 1.3726\pm 0.0047$  &748.86053 &  $ 1.1776\pm 0.0070$  \\ 
415.79370 & $ 1.3800\pm 0.0047$  &748.76599 & $ 1.3806\pm 0.0070$  &748.86255 &  $ 1.1566\pm 0.0070$  \\ 
415.79523 & $ 1.3729\pm 0.0047$  &748.76801 & $ 1.4026\pm 0.0070$  &748.86456 &  $ 1.1486\pm 0.0058$  \\ 
415.80130 & $ 1.3379\pm 0.0047$  &748.77002 & $ 1.4216\pm 0.0105$  &748.86658 &  $ 1.1326\pm 0.0047$  \\ 
415.80280 & $ 1.3299\pm 0.0047$  &748.77203 & $ 1.4186\pm 0.0082$  &748.86859 &  $ 1.1416\pm 0.0058$  \\ 
415.80890 & $ 1.2969\pm 0.0047$  &748.77606 & $ 1.4176\pm 0.0070$  &748.87061 &  $ 1.1116\pm 0.0047$  \\ 
415.81039 & $ 1.3060\pm 0.0070$  &748.77808 & $ 1.4286\pm 0.0070$  &748.87262 &  $ 1.1116\pm 0.0058$  \\ 
415.81647 & $ 1.2490\pm 0.0058$  &748.78009 & $ 1.4376\pm 0.0093$  &748.87463 &  $ 1.1116\pm 0.0058$  \\ 
415.81796 & $ 1.2450\pm 0.0058$  &748.78210 & $ 1.4346\pm 0.0093$  &748.87665 &  $ 1.0956\pm 0.0070$  \\ 
415.82407 & $ 1.2290\pm 0.0082$  &748.78412 & $ 1.4416\pm 0.0128$  &748.87866 &  $ 1.1236\pm 0.0303$  \\ 
415.82556 & $ 1.1930\pm 0.0070$  &748.78607 & $ 1.4506\pm 0.0082$  &748.88269 &  $ 1.0736\pm 0.0093$  \\ 
415.83167 & $ 1.1869\pm 0.0082$  &748.78809 & $ 1.4556\pm 0.0082$  &748.88470 &  $ 1.0566\pm 0.0093$  \\ 
415.85745 & $ 1.0680\pm 0.0070$  &748.79010 & $ 1.4556\pm 0.0093$  &748.88672 &  $ 1.0596\pm 0.0093$  \\ 
415.85898 & $ 1.0539\pm 0.0047$  &748.79211 & $ 1.4496\pm 0.0128$  &748.89075 &  $ 1.0496\pm 0.0070$  \\ 
415.86505 & $ 1.0350\pm 0.0070$  &748.79413 & $ 1.4426\pm 0.0105$  &748.89478 &  $ 1.0316\pm 0.0047$  \\ 
415.86655 & $ 1.0330\pm 0.0047$  &748.79816 & $ 1.4536\pm 0.0105$  &748.89679 &  $ 1.0416\pm 0.0047$  \\ 
415.87265 & $ 1.0170\pm 0.0047$  &748.80017 & $ 1.4416\pm 0.0128$  &748.89880 &  $ 1.0276\pm 0.0070$  \\ 
415.87415 & $ 1.0059\pm 0.0070$  &748.80219 & $ 1.4466\pm 0.0093$  &748.90082 &  $ 1.0336\pm 0.0058$  \\ 
415.88022 & $ 1.0079\pm 0.0047$  &748.80420 & $ 1.4596\pm 0.0082$  &748.90283 &  $ 1.0146\pm 0.0093$  \\ 
415.88174 & $ 1.0059\pm 0.0047$  &748.80621 & $ 1.4436\pm 0.0058$  &748.90485 &  $ 1.0206\pm 0.0093$  \\ 
415.88782 & $ 0.9990\pm 0.0070$  &748.80823 & $ 1.4306\pm 0.0058$  &748.90686 &  $ 1.0196\pm 0.0070$  \\ 
415.88934 & $ 1.0009\pm 0.0070$  &748.81024 & $ 1.4286\pm 0.0082$  &748.90881 &  $ 1.0136\pm 0.0047$  \\ 
415.89542 & $ 1.0039\pm 0.0058$  &748.81226 & $ 1.4136\pm 0.0047$  &748.91083 &  $ 1.0066\pm 0.0058$  \\ 
\tablebreak
415.89691 & $ 0.9990\pm 0.0070$  &748.81427 & $ 1.4106\pm 0.0082$  &748.91284 &  $ 1.0006\pm 0.0058$  \\ 
415.90302 & $ 0.9960\pm 0.0082$  &748.81628 & $ 1.4116\pm 0.0058$  &748.91486 &  $ 0.9906\pm 0.0058$  \\ 
415.90451 & $ 0.9950\pm 0.0047$  &748.81830 & $ 1.4076\pm 0.0070$  &748.91888 &  $ 1.0036\pm 0.0070$  \\ 
415.91061 & $ 1.0009\pm 0.0047$  &748.82031 & $ 1.3996\pm 0.0070$  &748.92090 &  $ 0.9956\pm 0.0070$  \\ 
415.91211 & $ 1.0000\pm 0.0047$  &748.82233 & $ 1.3816\pm 0.0070$  &748.92291 &  $ 1.0106\pm 0.0082$  \\ 
415.91821 & $ 0.9980\pm 0.0117$  &748.82635 & $ 1.3576\pm 0.0082$  &          &                       \\ 
415.91971 & $ 1.0020\pm 0.0082$  &748.82837 & $ 1.3456\pm 0.0058$  &          &                       \\ 
\enddata
\end{deluxetable}

\begin{deluxetable}{cc|cc|cc}
\tablecaption{KIC 6131659 $I$-band MLO Photometry\label{Idata}}
\tabletypesize{\scriptsize}
\tablewidth{0pt}
\tablehead{
\colhead{Time} &
\colhead{Instrumental $I$} &
\colhead{Time} &
\colhead{Instrumental $I$} &
\colhead{Time} &
\colhead{Instrumental $I$} \\
\colhead{(HJD-2,455,000)} &
\colhead{(mag)}  &
\colhead{(HJD-2,455,000)} &
\colhead{(mag)}  &
\colhead{(HJD-2,455,000)} &
\colhead{(mag)}  
}
\startdata
415.73608 & $ 1.3535\pm 0.0064$  &415.89386 & $ 1.0105\pm 0.0051$  &748.81079 &  $ 1.4026\pm 0.0116$  \\ 
415.73758 & $ 1.3505\pm 0.0077$  &415.89844 & $ 0.9895\pm 0.0116$  &748.81281 &  $ 1.3906\pm 0.0103$  \\ 
415.74365 & $ 1.3975\pm 0.0064$  &415.89996 & $ 1.0025\pm 0.0090$  &748.81482 &  $ 1.3866\pm 0.0103$  \\ 
415.74515 & $ 1.3875\pm 0.0051$  &415.90146 & $ 0.9965\pm 0.0103$  &748.81683 &  $ 1.3616\pm 0.0103$  \\ 
415.74664 & $ 1.3935\pm 0.0051$  &415.90604 & $ 1.0075\pm 0.0064$  &748.81885 &  $ 1.3606\pm 0.0090$  \\ 
415.75125 & $ 1.4185\pm 0.0051$  &415.90753 & $ 1.0005\pm 0.0051$  &748.82086 &  $ 1.3686\pm 0.0103$  \\ 
415.75424 & $ 1.4145\pm 0.0064$  &415.90906 & $ 1.0055\pm 0.0077$  &748.82288 &  $ 1.3496\pm 0.0103$  \\ 
415.75882 & $ 1.4285\pm 0.0064$  &415.91364 & $ 1.0005\pm 0.0077$  &748.82489 &  $ 1.3356\pm 0.0090$  \\ 
415.76031 & $ 1.4305\pm 0.0064$  &415.91513 & $ 1.0005\pm 0.0077$  &748.82690 &  $ 1.3466\pm 0.0116$  \\ 
415.76181 & $ 1.4305\pm 0.0064$  &415.91666 & $ 0.9995\pm 0.0064$  &748.82892 &  $ 1.3276\pm 0.0064$  \\ 
415.76639 & $ 1.4225\pm 0.0064$  &415.92123 & $ 1.0025\pm 0.0077$  &748.83087 &  $ 1.3156\pm 0.0077$  \\ 
415.76791 & $ 1.4325\pm 0.0077$  &415.92273 & $ 0.9985\pm 0.0090$  &748.83289 &  $ 1.2896\pm 0.0090$  \\ 
415.76941 & $ 1.4275\pm 0.0064$  &415.92422 & $ 1.0015\pm 0.0077$  &748.83490 &  $ 1.2876\pm 0.0090$  \\ 
415.77399 & $ 1.4165\pm 0.0051$  &415.92883 & $ 1.0085\pm 0.0064$  &748.83691 &  $ 1.2696\pm 0.0129$  \\ 
415.77548 & $ 1.4125\pm 0.0051$  &415.93033 & $ 1.0095\pm 0.0051$  &748.83893 &  $ 1.2756\pm 0.0090$  \\ 
415.77698 & $ 1.4065\pm 0.0064$  &415.93182 & $ 1.0095\pm 0.0064$  &748.84094 &  $ 1.2596\pm 0.0090$  \\ 
415.78159 & $ 1.4005\pm 0.0051$  &415.93640 & $ 0.9975\pm 0.0064$  &748.84296 &  $ 1.2546\pm 0.0077$  \\ 
415.78308 & $ 1.3835\pm 0.0077$  &415.93790 & $ 1.0055\pm 0.0103$  &748.84497 &  $ 1.2436\pm 0.0090$  \\ 
415.78458 & $ 1.3905\pm 0.0051$  &415.93942 & $ 1.0025\pm 0.0090$  &748.84698 &  $ 1.2296\pm 0.0090$  \\ 
415.78915 & $ 1.3785\pm 0.0064$  &415.94400 & $ 1.0045\pm 0.0090$  &748.84900 &  $ 1.2226\pm 0.0090$  \\ 
415.79065 & $ 1.3625\pm 0.0051$  &415.94550 & $ 1.0035\pm 0.0077$  &748.85101 &  $ 1.2186\pm 0.0090$  \\ 
415.79214 & $ 1.3745\pm 0.0051$  &415.94699 & $ 1.0025\pm 0.0077$  &748.85303 &  $ 1.1996\pm 0.0090$  \\ 
415.79672 & $ 1.3415\pm 0.0077$  &415.95160 & $ 0.9995\pm 0.0064$  &748.85504 &  $ 1.1906\pm 0.0116$  \\ 
415.79825 & $ 1.3435\pm 0.0090$  &415.95309 & $ 0.9915\pm 0.0051$  &748.85706 &  $ 1.1726\pm 0.0116$  \\ 
415.79974 & $ 1.3265\pm 0.0051$  &415.95459 & $ 0.9955\pm 0.0064$  &748.85907 &  $ 1.1776\pm 0.0103$  \\ 
415.80432 & $ 1.2965\pm 0.0090$  &415.95917 & $ 0.9875\pm 0.0064$  &748.86108 &  $ 1.1616\pm 0.0103$  \\ 
415.80582 & $ 1.2995\pm 0.0064$  &748.75446 & $ 1.3086\pm 0.0090$  &748.86310 &  $ 1.1526\pm 0.0116$  \\ 
415.80731 & $ 1.2875\pm 0.0090$  &748.75647 & $ 1.3236\pm 0.0077$  &748.86511 &  $ 1.1406\pm 0.0090$  \\ 
415.81192 & $ 1.2685\pm 0.0077$  &748.75848 & $ 1.3296\pm 0.0090$  &748.86713 &  $ 1.1396\pm 0.0077$  \\ 
415.81342 & $ 1.2525\pm 0.0077$  &748.76050 & $ 1.3316\pm 0.0090$  &748.86914 &  $ 1.1276\pm 0.0064$  \\ 
415.81491 & $ 1.2475\pm 0.0077$  &748.76251 & $ 1.3326\pm 0.0077$  &748.87115 &  $ 1.1196\pm 0.0103$  \\ 
415.81949 & $ 1.2245\pm 0.0090$  &748.76453 & $ 1.3556\pm 0.0077$  &748.87317 &  $ 1.1096\pm 0.0090$  \\ 
415.82098 & $ 1.2155\pm 0.0077$  &748.76654 & $ 1.3626\pm 0.0090$  &748.87518 &  $ 1.1076\pm 0.0090$  \\ 
415.82248 & $ 1.2035\pm 0.0077$  &748.76855 & $ 1.3746\pm 0.0129$  &748.87720 &  $ 1.0896\pm 0.0090$  \\ 
415.82709 & $ 1.1985\pm 0.0090$  &748.77057 & $ 1.3786\pm 0.0116$  &748.88324 &  $ 1.0766\pm 0.0103$  \\ 
415.82858 & $ 1.1735\pm 0.0103$  &748.77258 & $ 1.3786\pm 0.0129$  &748.88525 &  $ 1.0546\pm 0.0103$  \\ 
415.83008 & $ 1.1815\pm 0.0090$  &748.77460 & $ 1.3946\pm 0.0116$  &748.88721 &  $ 1.1096\pm 0.0180$  \\ 
415.83466 & $ 1.1455\pm 0.0077$  &748.77661 & $ 1.3996\pm 0.0116$  &748.88928 &  $ 1.0506\pm 0.0244$  \\ 
415.83618 & $ 1.1505\pm 0.0116$  &748.77856 & $ 1.3906\pm 0.0103$  &748.89124 &  $ 1.0516\pm 0.0090$  \\ 
415.83768 & $ 1.1315\pm 0.0116$  &748.78058 & $ 1.4026\pm 0.0116$  &748.89325 &  $ 1.0556\pm 0.0167$  \\ 
415.86047 & $ 1.0575\pm 0.0051$  &748.78259 & $ 1.4186\pm 0.0090$  &748.89526 &  $ 1.0576\pm 0.0077$  \\ 
415.86200 & $ 1.0465\pm 0.0051$  &748.78461 & $ 1.4086\pm 0.0116$  &748.89728 &  $ 1.0506\pm 0.0077$  \\ 
415.86349 & $ 1.0445\pm 0.0064$  &748.78662 & $ 1.4076\pm 0.0116$  &748.89929 &  $ 1.0276\pm 0.0090$  \\ 
415.86807 & $ 1.0275\pm 0.0064$  &748.78864 & $ 1.4286\pm 0.0129$  &748.90131 &  $ 1.0446\pm 0.0103$  \\ 
415.86957 & $ 1.0245\pm 0.0051$  &748.79065 & $ 1.4216\pm 0.0142$  &748.90332 &  $ 1.0276\pm 0.0090$  \\ 
415.87109 & $ 1.0215\pm 0.0051$  &748.79266 & $ 1.4176\pm 0.0116$  &748.90533 &  $ 1.0166\pm 0.0103$  \\ 
415.87567 & $ 1.0205\pm 0.0064$  &748.79468 & $ 1.4246\pm 0.0116$  &748.90735 &  $ 1.0056\pm 0.0116$  \\ 
415.87717 & $ 1.0105\pm 0.0077$  &748.79669 & $ 1.4166\pm 0.0129$  &748.90936 &  $ 1.0316\pm 0.0090$  \\ 
415.87866 & $ 1.0105\pm 0.0064$  &748.79871 & $ 1.4186\pm 0.0129$  &748.91138 &  $ 1.0166\pm 0.0090$  \\ 
415.88327 & $ 1.0035\pm 0.0077$  &748.80072 & $ 1.3946\pm 0.0154$  &748.91339 &  $ 1.0126\pm 0.0116$  \\ 
415.88477 & $ 1.0105\pm 0.0064$  &748.80273 & $ 1.4076\pm 0.0090$  &748.91541 &  $ 1.0016\pm 0.0129$  \\ 
415.88626 & $ 1.0095\pm 0.0064$  &748.80475 & $ 1.4106\pm 0.0090$  &748.91742 &  $ 1.0096\pm 0.0090$  \\ 
415.89087 & $ 1.0035\pm 0.0064$  &748.80676 & $ 1.4046\pm 0.0090$  &748.91943 &  $ 0.9916\pm 0.0103$  \\ 
415.89236 & $ 1.0125\pm 0.0051$  &748.80878 & $ 1.3986\pm 0.0090$  &748.92145 &  $ 1.0146\pm 0.0090$  \\ 
\enddata
\end{deluxetable}

\begin{deluxetable}{crrr}
\tablecaption{KIC 6131659 Radial Velocities From BF Analysis\label{RVdata}}
\tablewidth{0pt}
\tablehead{
\colhead{Time} &
\colhead{${\rm RV}_A$} &
\colhead{${\rm RV}_B$} &
\colhead{${\rm RV}_C$} \\
\colhead{(HJD-2,455,000)} &
\colhead{(km s$^{-1}$)} &
\colhead{(km s$^{-1}$)} &
\colhead{(km s$^{-1}$)} }
\startdata
426.81251  &  $ 38.44\pm 0.47$  &  $-31.01\pm 0.60$  &  $ 10.77\pm 0.69$   \\
498.61789  &  $ 47.52\pm 0.39$  &  $-47.24\pm 0.50$  &  $ 10.55\pm 0.54$   \\
499.63272  &  $ 48.27\pm 0.40$  &  $-45.31\pm 0.50$  &  $ 11.19\pm 0.51$   \\
505.59428  &  $-21.06\pm 0.43$  &  $ 47.01\pm 0.50$  &  $ 10.46\pm 0.77$   \\
524.55405  &  $-30.69\pm 0.50$  &  $ 61.86\pm 0.64$  &  $ 10.74\pm 0.98$   \\
631.02252  &  $-32.30\pm 0.59$  &  $ 62.08\pm 0.95$  &  $  9.22\pm 0.09$   \\
655.96600  &  $ 47.50\pm 0.45$  &  $-44.53\pm 0.56$  &  $ 12.04\pm 0.76$   \\
666.90528  &  $-28.16\pm 0.46$  &  $ 56.67\pm 0.61$  &  $ 11.72\pm 0.68$   \\
666.92570  &  $-28.03\pm 0.47$  &  $ 56.51\pm 0.61$  &  $  9.99\pm 0.89$   \\
673.92660  &  $ 48.79\pm 0.44$  &  $-46.10\pm 0.54$  &  $ 11.87\pm 0.68$   \\
709.78938  &  $ 48.68\pm 0.43$  &  $-46.05\pm 0.54$  &  $ 11.36\pm 0.64$   \\
718.78790  &  $-31.66\pm 0.49$  &  $ 61.85\pm 0.63$  &  $ 10.29\pm 0.80$   \\
752.70148  &  $-32.25\pm 0.42$  &  $ 62.55\pm 0.55$  &  $  9.63\pm 0.66$   \\
\enddata
\end{deluxetable}

\begin{deluxetable}{rr}
\tablecaption{KIC 6131659 Spectroscopic Parameters\label{specparm}}
\tablewidth{0pt}
\tablehead{
\colhead{parameter} &          
\colhead{value}         
}
\startdata
$P$ (days)  &  $17.5277 \pm 0.0031$ \\
$T_0$\tablenotemark{a}   &  $2,455,450.829\pm 0.023$ \\
$e$        &  0.0 (fixed) \\
$\omega$ (deg) &  \nodata  \\
$K_1$  (km s$^{-1}$) &  $40.91\pm 0.14$  \\
$K_2$  (km s$^{-1}$) &  $55.08\pm 0.18$  \\
$\gamma$ (km s$^{-1}$) & $8.142 \pm 0.008$ \\
$T_1$ (K)\tablenotemark{b}    & $5660\pm 140$ \\
$T_2$ (K)\tablenotemark{b}    & $4780\pm 105$ \\
$[m/H]$ (dex)\tablenotemark{b}  & $-0.23\pm 0.20 $
\enddata
\tablenotetext{a}{Time of primary eclipse.}
\tablenotetext{b}{Derived from TODCOR/TRICOR analysis.}
\end{deluxetable}


\begin{deluxetable}{rrrrrrrr}
\tabletypesize{\scriptsize}
\rotate
\tablecaption{Parameters from Light Curve Fitting\label{lightparm}}
\tablewidth{0pt}
\tablehead{
\colhead{} & 
\colhead{combined}         &
\colhead{short} &
\multicolumn{4}{c}{segments}  & 
\colhead{} \\
\colhead{parameter} &
\colhead{value} &
\colhead{cadence} &
\colhead{minimum} &
\colhead{maximum} &
\colhead{median} &
\colhead{rms} &
\colhead{adopted}
}
\startdata
$i$ (deg)     & $89.180\pm 0.001$      & $89.190\pm      0.003$
  & 89.112 & 89.208 & 89.186 & 0.019 &
  $89.186\pm 0.019$\\
$Q$           & $0.743\pm 0.008$    & $0.744\pm 0.007$    
  & 0.738 & 0.746 & 0.743 & 0.001  &
  $0.743\pm 0.008$  \\
$K_1$ (km s$^{-1}$)  & $40.92\pm 0.25$  & $40.95\pm 0.22$ 
   & 40.77  & 41.06  & 40.92  & 0.05 &
   $40.92\pm 0.25$   \\
$P$ (days)    & $17.52782716 \pm 0.00000017$  & \nodata
   &  \nodata & \nodata &\nodata & \nodata &
   $17.52782716(17)$  \\
$T_{\rm conj}$ (HJD 2,455,000+) & $450.820595\pm 0.000002$ &$293.070157\pm 0.000008$ 
   &  \nodata & \nodata &\nodata & \nodata &
   450.820595(2) \\
$e\cos\omega$\tablenotemark{a} ($\times 10^{-5}$)  & $6.94\pm 0.04$ & $6.88  \pm 0.02$
   & 6.61  & 7.26    & 6.98  & 0.20  &
   $6.88\pm 0.02$   \\
$e\cos\omega$\tablenotemark{b} ($\times 10^{-5}$)  & \nodata  & \nodata
   & \nodata  & \nodata    & \nodata  & \nodata  &
   $4.57\pm 0.02$   \\
$R_1/a$          & $0.026492\pm  0.000009$  & $0.026505 \pm 0.000023$
  &0.026296  &0.026660  &0.026460  & 0.000086 &
  $0.026460 \pm  0.000086$ \\
$R_2/a$          & $0.019223\pm  0.000009$ & $0.019104   \pm 0.000032$ 
  &0.018880  & 0.019589 & 0.019221  & 0.000182 &
   $0.019221  \pm  0.000182$ \\
$T_1$ (K)        & $5721\pm 5$  &  $5834\pm 20$  
  &5659    &5839    &5789    &50  &
   $5789\pm 50$ \\
$T_2/T_1$        & $0.79862\pm 0.00099$  & $ 0.79383  \pm     0.00072$
  & 0.79415 &0.79986 &0.79596 &0.00158 &
   $0.79596 \pm 0.00158$ \\
$T_2$  (K)       & $4569\pm 18$ & $4631\pm 20$ 
   & 4526 & 4642 & 4609 & 32 &
   $4609\pm 32$  \\
$T_3$  (K)       & $5258\pm 5$  & $4702\pm 60$
  & 4627 & 5371 & 4779 & 195 &
   $4779\pm 195$ \\
$S$              & $0.1700\pm0.0010$  & $0.330  \pm     0.028$
  &0.191 &0.364 &0.293 &0.051 &
  $0.293\pm 0.051$  \\
$k$              & $\le 0.001$ & $\le 0.0022$ 
  &0.00000 &0.00451 &0.00070 &0.00117 & 
  $\le 0.001$   \\ 
$x_1(K)$         & $0.517\pm 0.006$  & $0.469    \pm   0.020$
  &0.400 & 0.685 & 0.490 &0.059 & 
  $0.490\pm 0.059$  \\
$y_1(K)$         & $0.020\pm 0.010$  & $0.072   \pm    0.036$    
  & $-0.157$ & 0.269 & $0.050$ & 0.105 &
   $0.050\pm 0.105$  \\  
$x_2(K)$         & $0.703\pm 0.014$ & $0.441      \pm 0.047$   
  &0.365 & 0.938 & 0.623 & 0.137 &
  $0.623\pm 0.137$  \\
$y_2(K)$         & $-0.129\pm 0.022$ & $0.327\pm 0.080$    
  &$-0.497$  & 0.344 & $0.030$ & 0.214  & 
  $0.030\pm 0.214$  \\  
$x_1(V)$       & $0.630\pm0.317$  & $0.690\pm 0.308$
  & 0.487  &  0.939  & 0.787 & 0.133  &
  $0.630\pm 0.317$   \\ 
$y_1(V)$      &  $0.190\pm 0.528$ & $-0.012\pm 0.467$ 
  & $-0.353$  &  0.290 & $-0.099$ & 0.194  & 
  $0.190 \pm 0.520$  \\
$x_1(R)$         & $0.852\pm 0.098$  & $0.856  \pm     0.112$     
  &0.737 &0.944  & 0.866 & 0.042 &
   $0.852\pm 0.098$  \\
$y_1(R)$         & $-0.495\pm 0.290$ & $-0.480\pm 0.297$    
  & $-0.500$ & $-0.258$ & $-0.500$  & 0.062  &
    $-0.495\pm 0.290$  \\  
$x_1(I)$         & $0.510\pm 0.089$ & $0.626   \pm    0.124$   
  &0.512 &0.678 &0.602 &0.046 & 
  $0.510\pm 0.089$ \\
$y_1(I)$         & $-0.487\pm 0.215$ & $-0.496  \pm     0.283$  
  &$-0.500$ & $-0.322$&$ -0.500$ & 0.036 &
   $-0.487\pm 0.215$ \\  
\hline
$M_1$ ($M_{\odot}$)  & $0.922\pm 0.007$  & $0.924\pm 0.007$ 
  & 0.912  & 0.927   & 0.922  & 0.002  &
  $0.922\pm 0.007$ \\
$M_2$ ($M_{\odot}$)  & $0.685\pm 0.005$  & $0.686\pm 0.004$ 
  & 0.683  & 0.687  & 0.685  & 0.001 &
   $0.685\pm 0.005$   \\
$R_1$ ($R_{\odot}$)  & $0.8808\pm 0.0018$  & $0.8816\pm 0.0019$ 
  & 0.8743 & 0.8863 & 0.8800 & 0.0028  &
   $0.8800\pm 0.0028$  \\
$R_2$ ($R_{\odot}$)  & $0.6391\pm 0.0013$  & $0.6355\pm 0.0016$ 
  & 0.6277 & 0.6513 & 0.6395  & 0.0061 &
    $0.6395\pm 0.0061$   \\
$a$ ($R\,M_{\odot}$) & $33.247\pm 0.065$ & $33.264\pm 0.047$ 
   & 33.246 & 33.292 & 33.247 & 0.012 &
    $33.247\pm 0.065$ \\
\enddata
\tablenotetext{a}{Not corrected for light travel time.}
\tablenotetext{b}{Corrected for light travel time.}
\end{deluxetable}

\vspace{-10em}

\clearpage

\end{document}